\documentclass[11pt ]{article}
\usepackage{amsmath}
\usepackage{graphicx}
\usepackage{hyperref}
\usepackage{lmodern}
\usepackage{amssymb}
\usepackage{booktabs}
\usepackage{tikz}
\usetikzlibrary{patterns}
\renewcommand{\theequation}
{\arabic{section}.\arabic{equation}}

\makeatletter
\def\eqnarray{ \stepcounter{equation} \let\@currentlabel=\theequation
 \global\@eqnswtrue
 \global\@eqcnt\z@
 \tabskip\@centering
 \let\\=\@eqncr
 $$\halign to \displaywidth\bgroup\@eqnsel\hskip\@centering
 $\displaystyle\tabskip\z@{##}$&\global\@eqcnt\@ne
 \hfil$\displaystyle{{}##{}}$\hfil
 &\global\@eqcnt\tw@$\displaystyle\tabskip\z@{##}$\hfil
 \tabskip\@centering&\llap{##}\tabskip\z@\cr}
\makeatother

\makeatletter
\def\@arrayacol{\edef\@preamble{\@preamble \hskip .5\arraycolsep}}
\def\array{\let\@acol\@arrayacol \let\@classz\@arrayclassz
\let\@classiv\@arrayclassiv \let\\\@arraycr\def\@halignto{}\@tabarray}
\makeatother

\makeatletter
\newcounter{subeqncnt}
\def\thesubeqncnt{\alph{subeqncnt}}
\def\subequations{\begingroup%
   \stepcounter{equation}\edef\@tempa{\theequation}%
   \let\c@equation\c@subeqncnt\c@subeqncnt\z@
   \edef\theequation{\@tempa\noexpand\thesubeqncnt}}

\makeatother

\newcommand{\be}{\begin{equation}}
\newcommand{\ee}{\end{equation}}

\newcommand{\beqa}{\begin{eqnarray}}
\newcommand{\eeqa}{\end{eqnarray}}
\newcommand{\nn}{\nonumber}

\def\CB {{\cal B}}

\def\CD {{\cal D}}
\def\CE {{\cal E}}
\def\CF {{\cal F}}
\def\CG {{\cal G}}
\def\CH {{\cal H}}

\def\CL {{\cal L}}
\def\CM {{\cal M}}

\begin{document}

\setlength{\baselineskip}{7mm}
\begin{titlepage}
\begin{flushright}
{\tt NRCPS-HE-70-2022} \\
March, 2022
\end{flushright}

\vspace{1cm}

\begin{center}
{\it \Large 
Stability of Yang Mills Vacuum  State 
}

\vspace{1cm}

{ {George~Savvidy  }}\footnote{savvidy(AT)inp.demokritos.gr}

\vspace{1cm}

 {\it Institute of Nuclear and Particle Physics, NCSR Demokritos, GR-15310 Athens, Greece}\\
{\it  A.I. Alikhanyan National Science Laboratory, Yerevan, 0036, Armenia}\\
{\it Institut f\"ur Theoretische Physik,Universit\"at Leipzig, D-04109 Leipzig, Germany }

\end{center}

\vspace{1cm}

\begin{abstract}

We examine the phenomena of the chromomagnetic gluon condensation in the Yang-Mills theory and the problem of stability of the chromomagnetic vacuum fields.  The apparent instability of the chromomagnetic vacuum fields is a result of quadratic approximation.  The stability is restored when the nonlinear interaction of negative/unstable modes is taken into account in the case of chromomagnetic vacuum fields and the interaction of the zero modes in the case of (anti)self-dual covariantly-constant vacuum fields.   All these vacuum fields are stable and indicate that the Yang-Mills vacuum is  highly degenerate  quantum state.

\end{abstract}

\end{titlepage}

\section{\it Introduction}

In the earlier investigations of the chromomagnetic gluon condensation  \cite{Savvidy:1977as, Matinyan:1976mp, Batalin:1976uv, PhDTheses} it was realised  that consideration of the vacuum polarisation in the quadratic  approximation \cite{Batalin:1976uv,tHooft:1976snw} displays an apparent instability of the vacuum fields\footnote{Here, and afterwards, the phrase  "vacuum fields" refers to the gauge fields that are the solutions of the sourseless Yang-Mills equation.}  due to the negative/unstable modes  \cite{Nielsen:1978rm, Skalozub:1978fy, Ambjorn:1978ff, Nielsen:1978zg, Nielsen:1978nk, Ambjorn:1980ms,   Leutwyler:1980ev, Leutwyler:1980ma, Minkowski:1981ma, Flory:1983td, Faddeev:2001dda, Pimentel:2018nkl, Savvidy:2019grj}.  Our aim is to demonstrate that the stability is restored when the nonlinear interaction of negative/unstable modes is taken into account in the case of chromomagnetic vacuum fields and the interaction of the zero modes in the case of (anti)self-dual covariantly-constant vacuum fields.

We will consider first  the (anti)self-dual  covariantly-constant vacuum fields (\ref{selfdulitycon}) that have only positive/stable and infinitely many zero modes, so called chromons,  as it was advocated by Leutwyller  and Minkowski  \cite{ Leutwyler:1980ev, Leutwyler:1980ma, Minkowski:1981ma}, and, importantly, there are no negative/unstable modes. 
To calculate the contribution of infinite number of zero modes we suggested a  regularisation method that allows to sum the contribution of zero modes and get renormalised effective Lagrangian that has contribution of all positive/stable modes and zero modes.  In the second approach we are taking into account a nonlinear interaction of zero modes that provides a necessary convergence of the path integral and leads to the same result for the effective Lagrangian that does not contain an imaginary part {\cite{Savvidy:1977as}.

 Next we are considering the stability of general covariantly-constant  chromomagnetic vacuum fields (\ref{alpharegeul}).  Instead of zero modes, here appear a plethora of negative/unstable modes \cite{Nielsen:1978rm, Skalozub:1978fy, Ambjorn:1978ff, Nielsen:1978zg, Nielsen:1978nk, Ambjorn:1980ms}. Generalising  the calculation that was advocated earlier  by Ambjorn, Nielsen, Olesen \cite{ Ambjorn:1978ff, Nielsen:1978zg, Nielsen:1978nk, Ambjorn:1980ms},  Flory \cite{Flory:1983td}  and other authors \cite{ Leutwyler:1980ev, Leutwyler:1980ma,  Minkowski:1981ma,  Faddeev:2001dda,  Pimentel:2018nkl,  Savvidy:2019grj, parthasarathy,Kay1983,Dittrich:1983ej, Zwanziger:1982na, kumar, Kondo:2013aga, Cho:2004qf,Pak:2020izt,Pak:2020obo, Pak:2017skw,Pak:2020fkt} we performed the integration over nonlinearly  interacting negative/unstable modes  and obtained the effective Lagrangian that includes contributions of positive/stable and negative/unstable modes and demonstrated the stability of  covariantly-constant  chromomagnetic vacuum fields. This consideration reflects a well known fact that a magnetic field does not produce work  and cannot create particle pairs from the vacuum \cite{Savvidy:1977as},  opposite to what takes place in the case of an electric field \cite{Sauter:1931zz,  Euler:1935zz, Heisenberg:1935qt, Schwinger:1951nm}. All these vacuum fields are stable and indicate that the Yang-Mills vacuum is a highly degenerate  quantum state.

The rest of the article is devoted to the discussion of the chromomagnetic condensation and to the large $N$ behaviour of the effective Lagrangian in the case of $SU(N)$ group.  A number of Appendixes  are  devoted to the technical details and  the renormalisation group.

\section{\it Euclidean Path Integral }

The quantum-mechanical amplitudes can be formulated as a sum over physical space-time trajectories \cite{Feynman:1948ur,Feynman:1949zx}, as well as a sum over unphysical  "trajectories" in Euclidean space  \cite{Leutwyler:1980ev, Leutwyler:1980ma, Polyakov:1987ez, Dyson:1952tj, Lipatov:1976ny, tHooft,Coleman}. Here we will analyse the amplitudes and the effective Lagrangian in Yang-Mills theory by using the  Euclidean path integral representation  \cite{tHooft:1976snw, Coleman,Coleman:1973jx}. 

For a single particle the Euclidean path integral determines the matrix elements of the operator $\exp{(-H T)}$ ( \cite{Coleman},  see Appendix D):
\beqa\label{euclid0}
\langle \vec{x}^{'} \vert e^{- H T} \vert \vec{x} \rangle = 
\sum_n e^{- E_n T} \psi_n(\vec{x}^{'}) \psi^{*}_n(\vec{x})=    N \int^{ \vec{x}^{~'}}_{\vec{x}} \CD\vec{x}(\tau) e^{-S_E[x(\tau)]},
\eeqa 
where the left-hand side of  (\ref{euclid0}) is defined in terms of physical quantities, while the right-hand side in (\ref{euclid0}) is defined  in unphysical Euclidean space \cite{Coleman}.
A similar approach can be applied to the quantum  gauge field theory when the
physical states are described by gauge-invariant wave functionals  $ \psi_n[\vec{A}]= \langle \vec{A} \vert E_n \rangle$ invariant with respect to infinitesimal gauge transformations of the three-dimensional vector gauge field  $\vec{A}$ \cite{Leutwyler:1980ev,  Leutwyler:1980ma}:
\beqa\label{euclidfield}
\langle \vec{A}^{~'} \vert e^{- H_{YM} T} \vert \vec{A} \rangle = 
\sum_n e^{- E_n T} \psi_n[\vec{A}^{~'}] \psi^{*}_n[\vec{A}]=   N  \int^{ \vec{A}^{~'}}_{\vec{A}} \CD \vec{A}(\tau) e^{-S_E[A(\tau)]} \equiv N_f  e^{-\int  d^4 x \CL_{eff}(\vec{A}^{'}, \vec{A})},~~~
\eeqa
where $H_{YM}  \psi_n[\vec{A}] =E_n  \psi[\vec{A}]$. 
The left-hand side involves only quantities defined in a physical space-time  with Yang-Mills Hamiltonian    
\be
H_{YM}= {1\over 2} \int d^3 x \Big[ (\vec{E}^{a}_{i})^2 + (\vec{H}^{a}_{i})^2 \Big],
\ee
where  
\be
E^{a}_{i} = -i {\delta  \over \delta A^{a}_{i}} \psi(\vec{A}),~~~~~~H^{a}_{i} = \epsilon_{ijk} \{\partial_j A^{a}_{k}+{1\over 2} f^{abc} A^{b}_{j} A^{c}_{k}  \} \psi(\vec{A}).
\ee
While the right-hand side of the equation (\ref{euclidfield}) involves an integral over  the Euclidean field  $A^{a}_{i}( \vec{x}, \tau )$ with the proper boundary values 
\be\label{data}
A^{a}_{i}(\vec{x},0) = A^{a}_{i}(\vec{x}),~~~~~  A^{a}_{i}(\vec{x},T) = A^{'a}_{i}(\vec{x}).
\ee
The Euclidean action is given by
\be\label{action}
S_E= {1\over 4} \int d^4 x G^{a}_{\mu\nu}G^{a}_{\mu\nu},
\ee
where 
$$
G^{a}_{\mu\nu}(A)= \partial_{\mu} A^{a}_{\nu} - \partial_{\nu} A^{a}_{\mu} +g f^{abc} A^{b}_{\mu}A^{c}_{\nu}
$$  
and $x_0 \rightarrow -ix_0=-i \tau$,~$A_0 \rightarrow i A_0$.
The Euclidean fields cannot be interpreted directly in a physical space. In particular, a real electric field corresponds to an imaginary electric field in Euclidean  space $ \vec{E} \rightarrow  i \vec{E}$ \cite{Leutwyler:1980ev,Coleman}.    Nevertheless, the integral over  Euclidean fields interpolating between $A^{a}_{i}(\vec{x})$ and $A^{'a}_{i}(\vec{x})$ does represent the physical quantities: the energy spectrum,  the wave functions and the effective Lagrangian (\ref{euclidfield}) \cite{Leutwyler:1980ev, Leutwyler:1980ma,  Polyakov:1987ez, Dyson:1952tj, Lipatov:1976ny, tHooft, Coleman}. 
 
The path integral representation requires summation  over Euclidean fields $A^{a}_{\mu}(\vec{x},\tau)$ that interpolate between the fixed boundary data $A^{a}_{i}(\vec{x})$ and $A^{'a}_{i}(\vec{x})$. We will analyse the behaviour of vacuum fields  in the vicinity of a given interpolating field $B^a_{\mu}(x)$ that starts at $A^{a}_{i}(\vec{x})$ and ends at $A^{'a}_{i}(\vec{x})$:
\be
A^{a}_{\mu}(x) = B^a_{\mu}(x) + a^a_{\mu}(x).
\ee
Expanding the field strength  $G^a_{\mu\nu}(A)$  in powers of quantum field $a^a_{\mu}(x)$ that has the zero boundary values
\be
G^a_{\mu\nu}(A) = G^a_{\mu\nu}(B) + \nabla^{ab}_{\mu} a^{b}_{\nu}  - \nabla^{ab}_{\nu} a^{b}_{\mu}+ g f^{abc} a^{b}_{\mu}a^{c}_{\nu}, 
\ee
where $\nabla^{ab}_{\mu}$ is a covariant derivative with respect to the interpolating  field $B^a_{\mu}(x)$
\be
\nabla^{ab}_{\mu} a^{b}_{\nu} = \partial_{\mu} a^{a}_{\nu} +g f^{abc} B^{b}_{\mu}a^{c}_{\nu}
\ee
and $[\nabla_{\mu},\nabla_{\nu}]^{ab} = g f^{acb} G^{c}_{\mu\nu}(B)$. For the Euclidean action (\ref{action})  we will get:
\beqa\label{fluctuationall}
S_E&=&{1\over 4}\int d^4x G^a_{\mu\nu}(B) G^a_{\mu\nu}(B) + \int  d^4x G^a_{\mu\nu}(B) \nabla^{ab}_{\mu} a^{a}_{\nu}+\nn\\
&&+ {1\over 4}\int  d^4 x \Big((\nabla^{ab}_{\mu} a^{b}_{\nu} -\nabla^{ab}_{\nu} a^{b}_{\mu}+ g f^{abc} a^{b}_{\mu}a^{c}_{\nu})^2 + 2 g G^a_{\mu\nu}(B) f^{abc} a^{b}_{\mu}a^{c}_{\nu} \Big).
\eeqa
The first term in $S_E$ doesn't depend on the quantum field $a^a_{\mu}(x)$ and can be factorised  in the path integral (\ref{euclidfield}). The interpolating field $B^a_{\mu}$ is supposed to satisfy the Euclidean equation of motion:
\be\label{YMeq}
\nabla_{\mu} G_{\mu\nu}(B)=0, 
\ee
so that the second term in (\ref{fluctuationall}) that is linear in $a^a_{\mu}(x)$ vanishes and the effective Lagrangian is a gauge invariant functional \cite{Batalin:1976uv}.   The quadratic part of the action has the following form:
\be\label{linear}
K_E= {1\over 4}\int d^4x \Big[(\nabla^{ab}_{\mu} a^{b}_{\nu} -\nabla^{ab}_{\nu} a^{b}_{\mu} )^2 + 2 g G^a_{\mu\nu}(B) f^{abc} a^{b}_{\mu}a^{c}_{\nu} \Big].
\ee
The positive eigenvalues of the above quadratic form provide convergence of the path integral $\int e^{-K_E} \CD a_{\mu}$ and will be referred  as stable, while the negative eigenvalues produce  a divergency of the path integral in the directions of the corresponding eigenfunctions  and will be referred as unstable. The nonlinear terms in the Euclidean Lagrangian (\ref{fluctuationall})  are cubic and quartic in the quantum field $a^a_{\mu}(x)$:
\be\label{nonlinear}
V_E=\int d^4x \Big[- g f^{abc} a^{b}_{\nu}a^{c}_{\mu} \nabla^{ad}_{\mu} a^{d}_{\nu} + {g^2 \over 4} (f^{abc}a^{b}_{\mu}a^{c}_{\nu})^2\Big] .
\ee
The stability of the vacuum fields is directly connected with the stability of Euclidean field  $B^a_{\mu}$ that interpolates between the boundary vacuum fields $A^{a}_{i}(\vec{x})$ and $A^{'a}_{i}(\vec{x})$. The interpolating field may or may not be stable. An interpolating field is stable if the action associated with it is smaller than the action of all neighbouring fields that have  the same boundary values.  The interpolating field is unstable  if the action is exponentially diverging  in some directions of the Hilbert space.

It follows that in order for the one-loop approximation to make sense, the  amplitude $e^{-S_E}$, when it is taken in a quadratic approximation $e^{-K_E}$ (\ref{linear}), must decay in all directions of the quantum field $a^a_{\mu}(x)$ over which we are integrating, that is, all eigenvalues of the quadratic form $K_E$ must be  positive. If this is not the case and some of the eigenvalues are negative, the background field $B^a_{\mu}(x)$ displays an apparent instability in this quadratic approximation and stability should be reconsidered in the nonlinear regime by including the nonlinear interaction $V_E$ of the quantum field $a^a_{\mu}(x)$.

Let us specify the boundary fields $A^{a}_{i}(\vec{x})$ and $A^{'a}_{i}(\vec{x})$ at Euclidean time $\tau = 0$ and $\tau  = T$ to be equal  $A^{a}_{i}(\vec{x})=A^{'a}_{i}(\vec{x})$. We will consider boundary vacuum field $A^{a}_{i}(\vec{x})$ to be a {\it covariantly-constant} field \cite{Brown:1975bc, Duff:1975ue,  Batalin:1976uv, PhDTheses} and therefore the interpolating field should obey the Euclidean Euler-Lagrange equation 
(\ref{YMeq}). A suitable four-dimensional  gauge field $B^a_{\mu}(x)$ that provides a possible solution has the following form:
\be\label{interpol}
B^a_{\mu}(x) = -{1\over 2} F_{\mu\nu} x_{\nu} \delta^{a}_{3}.   
\ee
The Euclidean transformations rotate the vectors $E_{i }$ and $H_i$ by two independent rotations, and we may therefore transform these vectors into the direction of the z-axis \cite{Leutwyler:1980ev}:
\be\label{backfields}
F_{12}=H,~~~F_{30}= E,
\ee
and the gauge invariants are: $\CF_E= {1\over 4} G^a_{\mu\nu}G^a_{\mu\nu} ={H^2 +E^2\over 2}$ and the $\CG_E ={1\over 4} G^a_{\mu\nu} \tilde{G}^a_{\mu\nu} = H E$. In order to analyse the stability of the interpolating field we should initially consider the corresponding eigenvalue problem  that appears in the quadratic approximation (\ref{linear}) of the action (\ref{fluctuationall}) for the quantum field $a^a_{\mu}(x)$:
 \be\label{quadratic}
 -\nabla^{ab}_{\mu}( \nabla^{bc}_{\mu} a^{c}_{\nu} - \nabla^{bc}_{\nu} a^{c}_{\mu}) + g f^{abc} G^b_{\mu\nu} a^{c}_{\mu} = \lambda a^{a}_{\nu}.
 \ee
It is convenient to decompose the field $a_{\mu}$ into the neutral and charged components\footnote{We will consider the $SU(2)$ gauge group in the remaining part of the article. }:
\be\label{poten}
a^a_{\mu}= (a^3_{\mu}, a_{\mu}, a^{-}_{\mu}), ~~~~a_{\mu}=  a^{1}_{\mu} +i  a^{2}_{\mu} , 
~~~~a^-_{\mu}=   a^{1}_{\mu} -i  a^{2}_{\mu}.
\ee
For the neutral and charged components we will get: 
\beqa
 -\partial_{\mu} ( \partial_{\mu} a^{3}_{\nu} - \partial_{\nu} a^{3}_{\mu})  = \lambda a^{3}_{\nu},~~~~
 -\nabla_{\mu}( \nabla_{\mu} a_{\nu} - \nabla_{\nu} a_{\mu}) + i g F_{\mu\nu} a_{\mu} = \lambda a_{\nu},
\eeqa
where $\nabla_{\mu} = \partial_{\mu}  -  {1\over 2} i g F_{\mu\nu} x_{\nu}$. Taking into account that
\be
[\nabla_{\mu},\nabla_{\nu}] = ig F_{\mu\nu} \nn
\ee
and imposing the background gauge fixing condition on the quantum field \cite{Batalin:1976uv,PhDTheses,tHooft:1976snw}
\be
\nabla^{ab}_{\mu} a^{b}_{\mu} = 0
\ee
we will get for the charged components the following equation: 
\be\label{eigenva}
 -\nabla_{\mu} \nabla_{\mu} a_{\nu}  + 2 i g F_{\mu\nu} a_{\mu} = \lambda a_{\nu}~.
\ee
Taking the interpolating field (\ref{interpol}) in the z-direction (\ref{backfields}) we will get
\beqa\label{theh0}
H_0=-\nabla_{\mu} \nabla_{\mu} = -(  \partial_{\mu}  -  {1\over 2} i g F_{\mu\nu} x_{\nu})
( \partial_{\mu}  -  {1\over 2} i g F_{\mu\rho} x_{\rho})=\nn\\
=  -\partial^2_1- \partial^2_2 + ig H (  x_2 \partial_1 -  x_1 \partial_2) + {g^2 \over 4} H^2 ( x^2_2+  x^2_1)\nn\\
  - \partial^2_3- \partial^2_0 + ig E (  x_0 \partial_3 -  x_3 \partial_0) + {g^2 \over 4} E^2 ( x^2_3+  x^2_0) .
\eeqa
$H_0$ is a sum of isomorphic  oscillators in the $(1,2)$ and $(3,0)$ planes. Introducing the operators \cite{Leutwyler:1980ev, Leutwyler:1980ma}
\beqa\label{newoperators}
c_i = \partial_i + {g\over 2} H x_i,~~~~~c^+_i =  -\partial_i + {g\over 2} H x_i,~~~~~i=1,2\nn\\
d_j = \partial_j + {g\over 2} E x_j,~~~~~d^+_j =  -\partial_j + {g\over 2} E x_j,~~~~~j=3,0
\eeqa
one can find that (see Appendix A)
\beqa
H_0=(c^+_1 +i c^+_2)(c_1 - i c_2)+(d^+_3 +i d^+_0)(d_3 - i d_0)+ g H + g E .
\eeqa
The eigenstates of the operator $H_0$ therefore are: 
\be\label{excitedsta}
\psi_{n m }= (c^+_1 +i c^+_2)^{n} (d^+_3 +i d^+_0)^{m} \psi_{00} = (g H)^n(x_1+i x_2)^n (g E)^m(x_3+ix_0)^m  \psi_{00},
\ee
where
\beqa\label{lowest}
 \psi_{00}=e^{- {g H\over 4}(x^2_1 + x^2_2 )   } e^{ - {g E\over 4}( x^2_3  + x^2_0 )  }
\eeqa
and the corresponding eigenvalues $H_0 \psi_{n m } = \lambda_0 \psi_{n m}$ are: 
\be\label{H0spect}
\lambda_0 = (2n +1) gH + (2m +1) gE  .
\ee
All eigenstates have infinite degeneracy because the states 
\beqa\label{zeromodeswave}
~~~\psi_{ nm}(n_0, m_0)= (c^+_1-i c^+_2)^{n_0}(d^+_3 -i d^+_0)^{m_0} \psi_{nm}
\eeqa
have identical eigenvalues (\ref{H0spect}) and are indexed by two integers $n_0, m_0 =0,1,2...$ (see Appendix A for details).  Now we can turn to the investigation of the eigenstates  of the operator  for the charged field (\ref{poten})
\be\label{Hoperator}
H_{\mu\nu}=- g_{\mu\nu }\nabla_{\lambda} \nabla_{\lambda}    - 2 i g F_{\mu\nu},
\ee
that appears in the equation (\ref{eigenva}). They  are:
\be\label{chargefields}
b = a_1 +i a_2, ~~~b^- = a_1 - i a_2, ~~~ h= a_3 + i a_0,~~~ h^-= a_3 - i a_0.
\ee
The eigenvalues corresponding to the fields $b,b^-$  and $h,h^-$ of the operator $H_{\mu\nu}$
take the following form:
\beqa\label{spectr}
b^-:~~~~~\lambda_1=  (2n +1) gH + (2m +1) gE + 2 g H\nn\\
b~:~~~~~~\lambda_2 = (2n +1) gH + (2m +1) gE - 2 g H\nn\\
h^-:~~~~~\lambda_3 = (2n +1) gH + (2m +1) gE + 2 g E\nn\\
h~:~~~~~~\lambda_4 = (2n +1) gH + (2m +1) gE - 2 g E, 
\eeqa
where $n,m=0,1,2...$. For the conjugate field  $a^-_{\mu}$ one can find the identical eigenvalues $\lambda_i$, $i=5,...,8$. As one can see, the negative eigenvalues appear in $\lambda_2, \lambda_4$ if $E \neq  H$. It follows that only in the case of (anti)self-dual  interpolating field \cite{Leutwyler:1980ev, Leutwyler:1980ma}
\be\label{selfdulitycon}
E=   H
\ee 
there are no negative/unstable eigenvalues in the spectrum. It also follows that when $n=m=0$, there is an infinite number of zero eigenvalues $\lambda_2 =  \lambda_4 = 0$  that appear due to the high symmetry of the self-dual field $E = H$ and the degeneracy (\ref{zeromodeswave}).  These are  Leutwyller zero mode chromons and they are linear combinations of (\ref{zeromodeswave}) ($b^{-}= a_1 - i a_2=0, ~h^{-}=a_3 - i a_0=0$):
\beqa\label{zero}
a_1(\xi) =i a_2(\xi)=  \sum_{n_0,m_0} \xi_{n_0 m_0} ~(c^+_1-i c^+_2)^{n_0}(d^+_3 -i d^+_0)^{m_0} \psi_{00}\nn\\
a_3(\eta) =i a_0(\eta)=  \sum_{n_0,m_0} \eta_{n_0 m_0} ~(c^+_1-i c^+_2)^{n_0}(d^+_3 -i d^+_0)^{m_0} \psi_{00},
\eeqa
where the zero-mode amplitudes (collective variables)  $\xi_{n_0 m_0}$ and $\eta_{n_0 m_0}$ are arbitrary complex numbers\footnote{The zero mode fields modify the interpolating  field $B_{\mu} \rightarrow B_{\mu} +  a_{\mu}(\xi,\eta)$ in such a fashion that it remains self-dual even for arbitrary large zero-mode amplitudes \cite{Leutwyler:1980ev, Leutwyler:1980ma}.   The interpolating  field $ B_{\mu} $  that is the solution of the  YM equation of motion (\ref{YMeq}) and boundary values (\ref{selfdulitycon}) was fond by Minkowski  \cite{Minkowski:1981ma}.}.  For the zero modes the quadratic form  (\ref{linear})  vanishes $K_E=0$  and the stability of the  interpolating field  $B^a_{\mu}(x)$  is determined by the nonlinear interaction term  $e^{-V_E}$  (\ref{nonlinear}) of the actin $S_E$. 

 In his original article Leutwyler remarked: "It does not seem to be possible to evaluate the integral over all self-dual fields exactly." And  then: "Our motivation for restricting ourselves to small chromon amplitudes is of a technical nature: we do not know how to do better." Our aim is to calculate the contribution of the zero modes exactly. In order to calculate the contribution of zero modes we will suggest two alternative methods in the forthcoming sections. First we will develop the method of the infrared regularisation of zero modes.

 \section{\it Effective Lagrangian. Contribution of Positive/stable Modes  }

The Euclidean path integral (\ref{euclidfield}) for positive/stable eigenvalues of the quadratic form $K_E$ is defined through the determinants of the operators $H_{\mu\nu}$ and $H_0$ \cite{Batalin:1976uv,PhDTheses}:
\beqa\label{oneloopdef}
  \CL^{(1)}_{positive~modes} V_3 T = &&{1\over 2} \ln Det H - \ln Det H_0 = {1\over 2} \CD eg \sum_{n,m,i}   \ln  \lambda_i (n,m) - \CD eg \sum_{n,m}  \ln \lambda_0(n,m)=\nn\\
&=&-{1\over 2} \CD eg \int {ds \over s} e^{- \sum_{n,m,i}\lambda_i (n,m) s  }+ \CD eg \int {ds \over s} e^{- \sum_{n,m}\lambda_0 (n,m) s },
\eeqa
where $\CD eg= \Big({g H\over 2\pi}\Big)^2 V_3 T$ is the degeneracy of the eigenstates  (see Appendix B (\ref{degeneracy})). After substituting the eigenvalues (\ref{spectr}) we will get (see Appendix C)
\beqa\label{contofnonzeromodes}
  \CL^{(1)}_{positive~modes}  &=&  - {g^2 H^2 \over 4\pi^2} \int {ds \over s}  \sum^{\infty}_{n,m=0} \Big( e^{-2 g H (n +m +2) s} + e^{-2 g H (n +m ) s} -  e^{-2 g H (n +m +1) s} \Big)\nn\\
&=&- {1 \over 8 \pi^2} \int {ds \over s}   {g^2 H^2 \over  \sinh^2 g H s}. 
\eeqa
By using the renormalisation condition \cite{Savvidy:1977as,Matinyan:1976mp}
\be\label{normalosation1}
{\partial \CL \over \partial \CF_E }|_{ g^2 \CF_E =\mu^4} = 1,
\ee
where $\CF_E  = {1\over 4} G^a_{\mu\nu}G^a_{\mu\nu}= H^2 $ for  the contribution of the positive/stable modes, we will get:
\beqa\label{sgablezeromode}
  \CL^{(1)}_{positive~modes} 
&= &-{1 \over 8\pi^2} \int {ds \over s}  \Big[ {g^2 H^2 \over   \sinh^2 g H s} -{ g^2 H^2 \over   \sinh^2   \mu^2  s   }   + { g^2 H^2 \mu^2  s  \cosh  \mu^2  s   \over   \sinh^3   \mu^2  s  } \Big] = -{ g^2 H^2 \over 48 \pi^2}  \Big[  \ln {2 g^2 H^2 \over \mu^4} -1\Big ]. \nn\\
~~~~~~~~~~
\eeqa
The zero eigenvalues in (\ref{spectr}) and the corresponding eigenfunctions  (\ref{zero})  that appear when $n=m=0$ should be considered separately.  In quadratic approximation the zero-mode fields are not suppressed by the exponential factor $e^{- K_E }$ because  $K_E=0$ and the  zero modes amplitudes  can grow unboundly.  The quantum field $a^a_{\mu}(x)$ can deviate considerably from the self-dual field $ B_{\mu}(x) $ in the neighbourhood of which we are investigating. It merely means that the integration in these directions is not Gaussian \cite{tHooft:1976snw}. In order to handle the zero modes contribution one should replace the measure $\CD \vec{A}(\tau)$ in the Euclidean path integral (\ref{euclidfield}) by introducing the collective variables in the zero-mode directions  (\ref{zero}):
\be\label{zeromodemeasure}
\prod^{zero~modes}_{a, \mu} \CD a^{a}_{\mu}  =J \prod^{\CD eg}_{n_0,m_0=0} d\xi^{'}_{n_0 m_0} d\xi^{''}_{n_0 m_0}d\eta^{'}_{n_0 m_0} d\eta^{''}_{n_0 m_0}, 
\ee 
where $\xi_{n_0 m_0}=\xi^{'}_{n_0 m_0} +i \xi^{''}_{n_0 m_0}  $,  $\eta_{n_0 m_0}=\eta^{'}_{n_0 m_0} +i \eta^{''}_{n_0 m_0}  $ and and $J$\footnote{The Jacobian $J$ here is a gauge field strength independent  function.} is the  Jacobian factor\cite{tHooft:1976snw, Coleman:1973jx}.  As far as the quadratic part of the action $S_E$ does not suppress the zero mode fluctuations we have to consider the contribution of zero modes by taking into account the cubic and quartic terms $e^{-V_E}$ in the Euclidean action $S_E$ (\ref{nonlinear}). We will develop this approach in the forthcoming sections.  

We found it useful to develop first an alternative approach that allows to calculate the zero mode contribution to the path integral by suggesting an appropriate  infrared regularisation.  As we will see at the end of this investigation, both methods, the infrared regularisation and integration over the nonlinear interaction of zero modes, lead to  the same result (\ref{euclidlagra}).   

\section{\it Infrared Regularisation of Zero Modes} 
 
Instead of performing the integration over the zero modes collective variables  $(\xi,\eta )$ in the path integral (\ref{euclidfield}),  in this section  we will consider an appropriate infrared regularisation of the spectrum (\ref{spectr}).   For that we will infinitesimally elevate the zero modes by adding the  terms $\alpha g H$ and $\alpha g E$ to the zero eigenvalues spectrum (\ref{spectr}):
\beqa\label{spectr1}
\lambda_2 = (2n +1) gH + (2m +1) gE - 2 g H + \alpha g H  \nn\\
\lambda_4 = (2n +1) gH + (2m +1) gE - 2 g E + \alpha g E , 
\eeqa
where $\alpha$ is a dimensionless infinitesimal parameter.   {\it As we will see, the result does not depend on $\alpha$.} For the self-dual field $E=H$ the contribution of these modes $\lambda_{2,4}=(2n+2m+\alpha) gH $ at $n=m=0$ to the effective Lagrangian   will be 
\be\label{infracon}
\CL^{(1)}_{zero~modes}  = -2 {g^2 H^2 \over 4 \pi^2} \int {ds \over s}  e^{- \alpha g H s}.
\ee
Thus the contribution of positive  (\ref{contofnonzeromodes}) and zero modes  (\ref{infracon}) to the effective Lagrangian will take the following form:
\beqa
\CL^{(1)}_E  =  \CL^{(1)}_{positive~modes} + \CL^{(1)}_{zero~modes} 
&=& -{1 \over 2\pi^2} \int {ds \over s}  \Big[ {g^2 H^2 \over 4 \sinh^2 g H s} + g^2 H^2 e^{- \alpha g H s} \Big].
\eeqa
By using the renormalisation condition  (\ref{normalosation1})  for the  effective Lagrangian  we will get
\beqa\label{oneloop}
\CL^{(1)}_{E}  
= &-{1 \over 2\pi^2} \int {ds \over s}  \Big[ {g^2 H^2 \over 4 \sinh^2 g H s} -{ g^2 H^2 \over 4 \sinh^2   \mu^2  s   }   + { g^2 H^2 \mu^2  s  \cosh  \mu^2  s   \over 4 \sinh^3   \mu^2  s  } +  \nn \\
&+g^2 H^2 \Big( e^{- \alpha g H s} - e^{-  \alpha  \mu^2  s  } +   { \alpha  \mu^2 s \over 2  }~    e^{- \alpha   \mu^2 s   }  \Big)\Big]=\nn\\ 
&= -{ g^2 H^2 \over 48 \pi^2}  \Big[  \ln {2 g^2 H^2 \over \mu^4} -1\Big ]  +  {g^2 H^2 \over 4 \pi^2}  \Big[  \ln {2 g^2 H^2 \over \mu^4} -1\Big]. 
\eeqa
 As one can see from the above result, the Lagrangian does not depend on the infrared regularisation parameter $\alpha$ that was introduced earlier to elevate the zero mode degeneracy.  This is in a good agreement with the  independence of the effective Lagrangian on the other infrared regularisation parameter $\mu$ in (\ref{normalosation1}) \cite{Savvidy:1977as,Matinyan:1976mp}.  It follows that the effective Lagrangian will take the following form:  
\beqa\label{euclidlagra}
\CL_E  
=  H^2 + {11 g^2 H^2 \over 48 \pi^2}  \Big[  \ln {2 g^2 H^2 \over \mu^4} -1\Big].
\eeqa  
 The Euclidean (anti)self-dual interpolating field is stable and the corresponding boundary covariantly-constant  vacuum field is also stable against quantum fluctuations.  

There is a perfect consistency between the result obtained in Euclidean formulation and the one obtained in a physical  space-time \cite{Savvidy:1977as,Savvidy:2019grj}:
\be\label{minlagra}
\CL_{YM}  
=- \CF  - {11 g^2 \CF  \over 48 \pi^2}  \Big[  \ln {2 g^2 \CF \over \mu^4} -1\Big], ~~~~~~~~~~~\CF= {1\over 4} G^a_{\mu\nu}G^a_{\mu\nu} >  0. 
\ee
As it was already mentioned, the electric field in a physical space-time  $\vec{E}$  corresponds to an imaginary  electric field     in Euclidian space $ \vec{E} \rightarrow  i \vec{E}$,     therefore one can map the gauge invariant operators as
\be
 \CF   = {1\over 2} (\vec{H}^2 -\vec{E}^2)  ~~\Rightarrow ~~   {1\over 2} (\vec{H}^2 +\vec{E}^2)_E = \CF_E.
\ee
For the self-dual vacuum field $E=H$ we have $\CF_E = H^2$, and the effective Lagrangian (\ref{minlagra})  transforms  into   the Euclidean  effective Lagrangian (\ref{euclidlagra}).  Because the invariant $\CF_E$  is a positive quantity, it follows that in a physical space-time  the corresponding $\CF $ is also positive and corresponds to the chromomagnetic field \cite{Savvidy:1977as,Savvidy:2019grj}. A similar result takes place for the anti-self-dual field.

\section{\it Nonlinear Interaction of Zero Modes}

Here we will turn to the second approach taking into account  a nonlinear interaction of zero modes by replacing the measure $\CD \vec{A}(\tau)$ in the Euclidean path integral (\ref{euclidfield}) by using the collective variables in the zero-mode directions  (\ref{zeromodemeasure}):
\be
\prod^{zero~modes}_{a, \mu} \CD a^{a}_{\mu}  =J \prod^{\CD eg}_{n_0,m_0=0} d\xi^{'}_{n_0 m_0} d\xi^{''}_{n_0 m_0}d\eta^{'}_{n_0 m_0} d\eta^{''}_{n_0 m_0}.  
\ee 
 As far as the quadratic part (\ref{linear}) of the action $S_E$ vanishes and does not suppress the zero-mode fluctuations, it follows that one should take into account the contribution of the  zero-mode fields  through the cubic and quartic interactions  in the Euclidean action $S_E$ (\ref{nonlinear}).   The cubic self-interaction term (\ref{nonlinear}) vanishes in the zero mode directions (\ref{zero}):
\be\label{cubic1}
V^{(3)}_E=- g \epsilon^{abc} \int  d^4 x \ a^{b}_{\nu}a^{c}_{\mu} \nabla^{ad}_{\mu} a^{d}_{\nu}=0.
\ee
Only the quartic interaction term is not vanishing in the zero mode directions and has the following form:
\beqa\label{zeromodepartionfunc}
Z^{zero~modes}&=&N \int \exp{\Big[- {g^2 \over 4} \int  d^4 x  \Big( (a^{a}_{\mu} a^{a}_{\mu})^2
-(a^{a}_{\mu}a^{a}_{\nu})^2  \Big)     ~  \Big]} ~\prod^{zero~modes}_{a, \mu} \CD a^{a}_{\mu}  ,
\eeqa
where the integration is over all zero modes.  The above zero mode partition function (\ref{zeromodepartionfunc}), if expanded in the coupling constant $g^2$, will generate an infinite number of multi-loop diagrams, and it seems impossible to calculate them. Nevertheless the exact calculation of the zero mode partition function is possible because the field strength dependence within the path integral (\ref{zeromodepartionfunc}) can be factorised. 

   Let us first consider the lowest state (\ref{lowest})  when $n=m=n_0=m_0=0$ in (\ref{excitedsta}) and (\ref{zeromodeswave}).   The solution (\ref{zero}) corresponding to the lowest state with $n_0=m_0=0$ is:
\beqa\label{zeroditections}
&a^1_1 = \xi^{'} \psi_{00},~~&a^2_1 = \xi^{''} \psi_{00},~~~~a^3_1=0\nn\\
&a^1_2 = \xi^{''} \psi_{00},~~&a^2_2 = -\xi^{'} \psi_{00},~~a^3_2=0\nn\\
&a^1_3 = \eta^{'} \psi_{00},~~&a^2_3 = \eta^{''} \psi_{00},~~~~a^3_3=0\nn\\
&a^1_0 = \eta^{''} \psi_{00},~~&a^2_0 = -\eta^{'} \psi_{00},~~a^3_0=0,
\eeqa
where $\xi_{00} = \xi^{'}  +i  \xi^{''}  $ and $\eta_{00} = \eta^{'}  +i  \eta^{''}  $.  For that lowest zero mode  field (\ref{zero}), (\ref{zeroditections}) the  quartic term will take the following form (see Appendix B for details):
\be\label{quartic}
V^{(4)}_E= {g^2 \over 4} \int d^4x (\epsilon^{abc}a^{b}_{\mu}a^{c}_{\nu})^2
=  {g^2 \over 2}(\xi^{' 2}+\xi^{'' 2}+\eta^{' 2}+\eta^{'' 2})^2 \int d^4x  \vert \psi_{00}(x) \vert^4,
\ee
and the corresponding part of the partition function can be represented in the following form:
\beqa
Z^{zero~mode}_{00}&=& \mu^4 \int \exp{\Big[- {g^2 \over 2}   (\xi^{' 2}+\xi^{'' 2}+\eta^{' 2}+\eta^{'' 2})^2   \int  d^4 x    \vert \psi_{00} \vert^4 ~   \Big] }  d \xi^{' } d \xi^{'' } d \eta^{' } d \eta^{'' } \nn\\
&=& \mu^4 \int \exp{\Big[- {g^2 \over 2}  \Big({g H  \over 4 \pi}\Big)^2   (\xi^{' 2}+\xi^{'' 2}+\eta^{' 2}+\eta^{'' 2})^2   \Big] }~ d \xi^{' } d \xi^{'' } d \eta^{' } d \eta^{'' },
\eeqa 
where we used the expression (\ref{normalisa1}).   Introducing the dimensionless variables $(\xi,\eta) \rightarrow (\xi,\eta)/(gH)^{1/2}$ allows to factorise the field strength dependence in the path integral, and we will get
\beqa
Z^{zero~mode}_{00}&=&   \Big({ \mu^2 \over g H } \Big)^2     \int \exp{\Big[- {g^2 \over 32}   (\xi^{' 2}+\xi^{'' 2}+\eta^{' 2}+\eta^{'' 2})^2   \Big] }~ d \xi^{' } d \xi^{'' } d \eta^{' } d \eta^{'' }= N_{00} \Big({ \mu^2 \over g H } \Big)^2.~~~~
\eeqa 
For the general zero mode field $\psi_{00}(n_0, m_0)  $ (\ref{zeromodeswave}) we will have  (see Appendix B (\ref{generalquartic2}))
\beqa\label{individualzero}
Z^{zero~mode}_{n_0,m_0}&=& \Big({ \mu^2 \over g H } \Big)^2     \int \exp{\Big[- {g^2 \over 32}  {\Gamma(n_0 +1/2) \Gamma(m_0 +1/2) \over \pi \Gamma(n_0+1) \Gamma(m_0+1)}  (\vert \xi_{n_0 m_0} \vert^2 +\vert \eta_{n_0 m_0} \vert^2)^2   \Big] }~ d^2 \xi_{n_0m_0} d^2 \eta_{n_0m_0}\nn\\
&=& N_{n_0,m_0} \Big({ \mu^2 \over g H } \Big)^2.     
\eeqa
In order to calculate the contributions of all individual self-interacting zero-modes  one should evaluate  the product $J \prod^{\CD eg}_{n_0=0,m_0=0} Z^{zero~mode}_{n_0,m_0}$ taking into account the degeneracy of the zero modes (\ref{degeneracy}):
\beqa\label{zeromodepro1}
Z_{zero~modes}&=& J \prod^{\CD eg}_{n_0=0,m_0=0} N_{n_0,m_0}  \Big( {  \mu^4 \over g^2 H^2 } \Big)   =
 N  \Big(  { \mu^4 \over g^2 H^2 }  \Big)^{ \CD eg} =N  e^{  -   {g^2 H^2 \over 4 \pi^2 }   \ln   { g^2 H^2  \over  \mu^4 }   V_3 T      } ,
\eeqa
where $N= J \prod^{\CD eg}_{n_0=0,m_0=0} N_{n_0,m_0}  $.  In the limit  $T \rightarrow \infty$  we have  $Z^{}_{zero~modes}  \rightarrow N e^{-\CL_{eff} V_3 T} $  and for the zero mode contribution to the effective Lagrangian  we will get
\be\label{zeromodenonlinear1}
\CL_{self-interacting~zero~modes} =  {g^2 H^2 \over 4 \pi^2 }    \ln   {  g^2 H^2  \over  \mu^4 }   .
\ee
This expression coincides  with our previous result (\ref{infracon}), (\ref{oneloop})  obtained by using the infrared regularisation. Now, adding the contribution of positive/stable  modes (\ref{sgablezeromode}) to the  (\ref{zeromodenonlinear1}) for the effective Lagrangian  we will get the expression that  coincides with the one obtained previously (\ref{euclidlagra}). 

The above result (\ref{zeromodepro1}) pointed out to the fact that exact integration over the zero modes  (\ref{zero}) can be performed not only for self-interacting but  also for fully interacting zero modes\footnote{I would like to thank Konstantin Savvidy for pointing out to me the possibility of such generalisation.}. The full interaction term has the following form (to be compared  with (\ref{quarticgen})):
\be\label{quarticgen1}
V_E ={g^2 \over 4} \int d^4x  (\epsilon^{abc}a^{b}_{\mu}a^{c}_{\nu})^2=  {g^2 \over 2}\int d^4x \Big[\Big\vert \sum_{n_0, m_0}   \xi_{n_0 m_0} \psi_{00}(n_0,m_0;x) \Big\vert^2 +\Big\vert \sum_{n_0 ,m_0} \eta_{n_0 m_0}\psi_{00}(n_0,m_0;x) \Big\vert^2 \Big]^2. 
\ee
The corresponding partition function will take the following form:
\beqa\label{intzeromodepath}
&Z_{zero~modes}= \nn\\
&=    \int \exp{\Big\{- {g^2 (g H)^2 \over 2}\int \Big( \Big\vert \sum_{n_0, m_0}   \xi_{n_0 m_0} \psi_{00}(n_0,m_0;y)  \Big\vert^2 
+\Big\vert \sum_{n_0, m_0} \eta_{n_0 m_0} \psi_{00}(n_0,m_0;y)\Big\vert^2 \Big)^2 ~d^4 y ~\Big\} }~  \nn\\
&J \prod^{\CD eg}_{n_0,m_0} \mu^4  d\xi^{'}_{n_0 m_0} d\xi^{''}_{n_0 m_0}d\eta^{'}_{n_0 m_0} d\eta^{''}_{n_0 m_0}, 
\eeqa 
where  we introduced the dimensionless variables $y_i = x_i  (g H )^{1/2}$ in the integral $V_E$, which allows to factorise the field strength dependence in the exponent  of the path integral (\ref{intzeromodepath}). The wave functions $\psi_{00}(n_0,m_0;x)$ were normalised as in (\ref{generalquartic2}), and in terms of the dimensionless variables they are:
$$
\psi_{00}(n_0,m_0;y)=  {(y_1-iy_2)^{n_0}  (y_3-i y_0)^{m_0} \over  ( \pi^2  2^{n_0+m_0 +2}n_0!  m_0!)^{1/2}   } \exp{\Big\{-{y^2_1 + y^2_2+ y^2_3  + y^2_0  \over 4}\Big\} }.
$$
By using the dimensionless collective variables $(\xi,\eta)  \rightarrow  (\xi,\eta) /(gH)^{1/2} $ introduced  above (\ref{individualzero}) we get 
\beqa
&Z_{zero~modes}=  \int \exp{\Big\{- {g^2  \over 2}\int \Big( \Big\vert \sum_{n_0, m_0} \xi_{n_0 m_0}  \psi_{n_0 m_0}(y) \Big\vert^2 
+\Big\vert \sum_{n_0, m_0} \eta_{n_0 m_0}   \psi_{n_0 m_0}(y) \Big\vert^2 \Big)^2 ~d^4 y ~\Big\} }~  \nn\\
&J \prod^{\CD eg}_{n_0,m_0} \Big({\mu^2 \over g H}\Big)^2 d\xi^{'}_{n_0 m_0} d\xi^{''}_{n_0 m_0}d\eta^{'}_{n_0 m_0} d\eta^{''}_{n_0 m_0}
\eeqa 
and  the field strength dependence completely factorises as well and  we have: 
\beqa\label{fullzeromodecont}
Z_{zero~modes}= N \prod^{\CD eg}_{n_0,m_0}  \Big( {  \mu^4 \over g^2 H^2 } \Big)   =
 N  \Big(  { \mu^4 \over g^2 H^2 }  \Big)^{ \CD eg} = N  e^{  -   {g^2 H^2 \over 4 \pi^2 }   \ln   { g^2 H^2  \over  \mu^4 }      V_3 T       },
\eeqa 
 where 
 \beqa
& N= \int \exp{\Big\{- {g^2  \over 2}\int \Big( \Big\vert \sum_{n_0, m_0} \xi_{n_0 m_0}  \psi_{n_0 m_0}(y) \Big\vert^2 
+\Big\vert \sum_{n_0, m_0} \eta_{n_0 m_0}   \psi_{n_0 m_0}(y) \Big\vert^2 \Big)^2 ~d^4 y ~\Big\} }~  \nn\\
&J \prod^{\CD eg}_{n_0,m_0}  d\xi^{'}_{n_0 m_0} d\xi^{''}_{n_0 m_0}d\eta^{'}_{n_0 m_0} d\eta^{''}_{n_0 m_0}. 
 \eeqa
In the limit  $T \rightarrow \infty$ from (\ref{fullzeromodecont})  for the zero mode effective Lagrangian   we will get
\be\label{zeromodenonlinear}
\CL_{interacting~zero~modes} =  {g^2 H^2 \over 4 \pi^2 }    \ln   {  g^2 H^2  \over  \mu^4 }   .
\ee
The effective Lagrangian is a sum of positive/stable (\ref{sgablezeromode}) and fully interacting zero modes (\ref{zeromodenonlinear}):  
\be
\CL^{eff}_E  =  H^2 + \CL^{(1)}_{positive~modes} + \CL_{interacting~zero~modes} =H^2 + {11 g^2 H^2 \over 48 \pi^2}  \Big[  \ln {2 g^2 H^2 \over \mu^4} -1\Big],
\ee
and it coincides with our previous results (\ref{euclidlagra}) and (\ref{zeromodenonlinear1}).  

It is remarkable that the zero mode contribution that was calculated in terms of  infrared regularisation of the spectrum (\ref{spectr1}), (\ref{infracon}), (\ref{oneloop}),  by integration over the self-interacting zero modes in (\ref{zeromodepro1}) and then by integration over fully interacting zero modes in (\ref{fullzeromodecont}), all lead to the same result indicating the robustness of the logarithmic structure of the effective Lagrangian  and that it is without an imaginary part \cite{Savvidy:1977as}. It is interesting to investigate  to what extent this behavioural robustness is rooted in the entropy factor,  through the degeneracy of the positive and zero mode states  and through the scaling invariance of the Yang-Mills action.

 \section{\it  Deformation of  (Anti)Self-Dual Field }

Breaking the (anti)self-duality condition (\ref{selfdulitycon}) will create negative/unstable modes (\ref{spectr})  if  we are considering the stability problem of the vacuum fields in the quadratic approximation \cite{Leutwyler:1980ev, Leutwyler:1980ma}.  The question is if the interaction of negative/unstable modes  will improve the stability of general covariantly-constant chromomagnetic vacuum fields  $H > E $.  As we will show below, even in that case, when there is a plethora of negative/unstable modes in the quadratic approximation,  the vacuum fields turn out to be stable due to the nonlinear interaction of negative/unstable modes. Here we will generalise the calculation that were advocated earlier  by Ambjorn, Nielsen, Olesen \cite{ Ambjorn:1978ff, Nielsen:1978zg, Nielsen:1978nk, Ambjorn:1980ms},  Flory \cite{Flory:1983td}  and other authors \cite{ Leutwyler:1980ev, Leutwyler:1980ma, Flory:1983td, Faddeev:2001dda,  Pimentel:2018nkl,   Savvidy:2019grj, parthasarathy,Kay1983,Dittrich:1983ej, Zwanziger:1982na, kumar, Kondo:2013aga, Cho:2004qf,Pak:2020izt,Pak:2020obo, Pak:2017skw}   by taking into account the interaction of negative/unstable modes.

 Let us perform  a deformation of the self-duality condition (\ref{selfdulitycon})  of the interpolating field $B^a_{\mu}(x)$ (\ref{backfields}) in the following manner \cite{Flory:1983td}:
\be\label{alpharegeul}
F_{12} = H,~~~~~F_{30}= E= \gamma H,~~~~~~\gamma \leq 1 ,
\ee
where $\gamma$ is a real parameter  that breaks the self-duality condition (\ref{selfdulitycon}). At $\gamma=1$ we will have the self-dual interpolating field (\ref{selfdulitycon}). The consequences of this deformation are two-fold: First of all, there will be positive/stable modes in the spectrum when $\gamma \neq 1$ and, secondly, instead of zero modes  the negative/unstable modes will appear in the spectrum.  The problem of integration over a nonlinearly interacting zero modes now turns into the problem of integration over a nonlinearly interacting negative/unstable modes. After the substitution $E \rightarrow \gamma H$ in (\ref{spectr}) the spectrum will take the following form:
\beqa\label{spectrbreack}
b^-:~~~~~\lambda_1=  (2n +1) gH + (2m +1) \gamma gH + 2 g H\nn\\
b~:~~~~~~\lambda_2 = (2n +1) gH + (2m +1) \gamma gH - 2 g H\nn\\
h^-:~~~~~\lambda_3 = (2n +1) gH + (2m +1) \gamma gH + 2 \gamma g H\nn\\
h~:~~~~~~\lambda_4 = (2n +1) gH + (2m +1) \gamma gH - 2 \gamma g H, 
\eeqa
where the eigenvalues $\lambda_2$ are negative  when $n=0$ and $0 \leq m \leq {1- \gamma \over 2 \gamma} $:
\be\label{spectrbreack1}
\lambda_2\vert_{n=0} =(2m \gamma -1 + \gamma) g H.
\ee
When $\gamma$ is a small number close to zero, the amount of negative/unstable modes is large, and when  $\gamma$ tends to one, the number of negative/unstable modes decreases, and they completely disappear at $\gamma=1$. At $\gamma=1$   instead of negative/unstable modes we will have the zero modes   $\lambda_2$ and $\lambda_4$ (\ref{spectr}),   the case of the (ani)self-dual interpolating field that we already considered in the previous sections.  

As we argued above, it is crucial  to consider  the nonlinear interaction of negative/unstable modes.  Let us first calculate the contribution of the positive/stable modes. Substituting the eigenvalues (\ref{spectrbreack}) into the (\ref{oneloopdef}) one can get: 
\beqa\label{contofnonzeromodes1}
  \CL^{(1)}_{positive~modes} &&=  - { \gamma g^2 H^2 \over 4\pi^2} \int {ds \over s}  \sum^{\infty}_{n,m=0} \Big( e^{-  g H (2 n +2  \gamma m +3+ \gamma) s} +  e^{- g H (2 n +2 \alpha m +1+3 \gamma) s} + \nn\\
&&+ e^{-  g H (2n +2\gamma m +1- \gamma) s} 
+ e^{- g H (2 n +2  \gamma m -1 +\gamma) s} - 
2 e^{- g H (2 n +2  \gamma m +1+ \gamma) s}   \Big) + \nn\\
&& + { \gamma g^2 H^2 \over 4\pi^2} \int {ds \over s} \sum^{{1\over 2 \gamma} -  {1\over 2}  } _{m=0}e^{- g H ( 2  \gamma m -1 + \gamma) s} .
 \eeqa
By performing the summation over $n$ and $m$ while using formulas in Appendix C one can get:
 \beqa\label{contofnonzeromodes2}
  \CL^{(1)}_{positive~modes} &=& -{\gamma g^2 H^2  \over 8 \pi^2} \int {ds \over s} \Big[  {1  \over  \sinh(g H s)    \sinh(g \gamma H s) } +2 {\sinh(g H s)    \over    \sinh(g \gamma H s) } +2 { \sinh(g \gamma H s)  \over  \sinh(g H s)    } \Big] +\nn\\
&&\nn\\
 &+& {\gamma g^2 H^2 \over 4\pi^2} \int {ds \over s}  e^{g H s(1-\gamma)} {1- e^{-g H s(1+\gamma)}  \over 1- e^{-2 g \gamma H s  }}.
\eeqa 
This expression represents a converging integral in the infrared region ($s \rightarrow \infty$) and can be renormalised in the ultraviolet region ($s \rightarrow 0$) by using the renormalisation condition (\ref{normalosation1}).  In the case of pure chromomagnetic field $\gamma = 0$ we will get: 
\beqa\label{gammazero}
\CL^{(1)}_{positive~modes}&=& -{1 \over 8 \pi^2} \int^{\infty}_{\infty} {ds \over s^3} \Big[  { g H s   \over  \sinh g H s    } -1 -{g^2 H^2 s\over 2 \mu^2}\Big({1\over \sinh \mu^2 s }    - {\mu^2 s \cosh{\mu^2 s}\over \sinh^2 \mu^2 s } \Big) - \nn\\
&-& g H s (e^{- g H s}-1) + {g^2  H^2 s\over 2 \mu^2} ( e^{-\mu^2 s} -1 -  \mu^2 s\ e^{-\mu^2 s} )\Big]=\nn\\
&=&-{g^2 H^2 \over 48 \pi^2} \Big( \ln   { g H  \over  \mu^2 }   - {1\over 2}\Big) +  {g^2 H^2 \over 8 \pi^2} \Big( \ln   { g H  \over  \mu^2 }   - {1\over 2}\Big),
\eeqa 
and in the case of  (anti)self-dual  field $\gamma =1$  the   $\CL^{(1)}_{positive~modes}$ (\ref{contofnonzeromodes2}) reduces to the expression   (\ref{sgablezeromode}).  Our main concern now is the contribution of the negative/unstable modes. The corresponding wave functions (\ref{excitedsta}) are: 
\be\label{setofunsmodes}
\psi_{0 m} =  \zeta_m (d^+_3 +i d^+_0)^{m} \psi_{00},~~~~~~~~n=0,~~~   0 \leq m \leq {1\over 2 \gamma} -  {1\over 2  },
\ee
and the corresponding field components (\ref{chargefields}) are:
\be\label{unstablemodes}
b=a_1+i a_2 = \zeta_m (d^+_3 +i d^+_0)^{m} \psi_{00},~~~b^-= a_1-i a_2 =0,
\ee
where $\zeta_m$ are the amplitudes of the negative/unstable modes and $h=h^- =0$. For the lowest state $m=0$  we have $\zeta_0 \equiv \zeta^{'}+i \zeta^{''}$:
\beqa\label{unstablemodes1}
&a^1_1 = \zeta^{'} \psi_{00},~~&a^2_1 = \zeta^{''} \psi_{00},~~~~a^3_1=0\nn\\
&a^1_2 = \zeta^{''} \psi_{00},~~&a^2_2 = -\zeta^{'} \psi_{00},~~a^3_2=0\nn\\
&a^1_3 =0,~~&a^2_3 = 0,~~~~~~~~~~a^3_3=0\nn\\
&a^1_0 = 0,~~&a^2_0 =0,~~~~~~~~~~a^3_0=0.
\eeqa 
The  contribution of the negative/unstable modes will take the following form: 
\beqa\label{uneqnon}
&Z_{negative~modes} =\nn\\
&= \int  \exp{\Big[-\int d^4 x\Big(~{1\over 2} ~a^{a}_{\mu}  (  -\delta_{\mu\nu}   \nabla^{ab}_{\rho} \nabla^{bc}_{\rho}     - 2 g f^{abc} G^b_{\mu\nu} ) a^{c}_{\nu}      + {g^2 \over 4}((a^{a}_{\mu} a^{a}_{\mu})^2
-(a^{a}_{\mu}a^{a}_{\nu})^2 ) \Big)       \Big] } \prod^{negative~modes }_{a \mu} \CD a^{a}_{\mu},  \nn\\
\eeqa
where the integration is over all negative/unstable modes:
\be\label{negativemeasure}
\prod^{negative~modes }_{a \mu} \CD a^{a}_{\mu}  = J \prod^{\CD eg}_{n_0,m_0}   \prod^{{1\over 2  \gamma} -  {1\over 2}  }_{m=0}  d \zeta^{'}_m  d \zeta^{''}_m .
\ee
The computation of the above negative mode partition function seems impossible if one ignores the scaling invariance of the Yang-Mills action and the degeneracy of the quantum states. At first we will consider the integration only over self-interacting negative/unstable modes. 

The cubic self-interaction term (\ref{cubic1}) for the negative/unstable modes (\ref{unstablemodes}), (\ref{unstablemodes}) vanishes. 
For the negative/unstable modes the quartic term is
\be\label{quarticgenunstable}
V_E =  {g^2 \over 2} \vert \zeta_m  \vert^4 ~\int d^4 x \vert \psi_{0m}(x) \vert^4
\ee
and the four-dimensional Euclidean integral is 
\be
{1\over C^2_{0} } {1\over C^2_{m} } \int dx_1 dx_2 d x_3 d x_0 \vert \psi_{0m}(x) \vert^4 = \gamma\Big({ g H  \over 4 \pi}\Big)^2  { \Gamma(m+1/2) \over \pi^{{1\over 2}} \Gamma(m+1)}.\nn
\ee
By substituting the negative/unstable modes (\ref{spectrbreack1}), (\ref{unstablemodes}) into the (\ref{uneqnon}) one can get:
\beqa
&Z_{negative~modes}^{self-interacting} = \mu^2 \prod^{{1\over 2  \gamma} -  {1\over 2}  }_{m=0}  \int  \exp{\Big[  g H ( 1-\gamma  - 2 m \gamma  )  \vert \zeta_m \vert^2  - {\gamma g^4 H^2 \over 32 \pi^2} { \Gamma(m+1/2) \over \pi^{{1\over 2}} \Gamma(m+1)}\vert \zeta_m \vert^4         \Big] } d \zeta^{'}_m  d \zeta^{''}_m ,~~\nn\\
\eeqa
where the product is over all negative/unstable modes (\ref{setofunsmodes}).
Introducing dimensionless collective variables $ \zeta_{m}  \rightarrow   \zeta_{m} /(gH)^{1/2} $ one can factorise the field dependence from each  integral in the product, and we will get  
\beqa
Z_{negative~modes}^{self-interacting} = \prod^{{1\over 2  \gamma} -  {1\over 2}  }_{m=0} \Big({\mu^2 \over g H} \Big)  \int  \exp{\Big[ ( 1-\gamma - 2 m \gamma  )  \vert \zeta_m \vert^2 -  {\gamma g^4   \over 32 \pi^2} { \Gamma(m+1/2) \over \pi^{{1\over 2}} \Gamma(m+1)}\vert \zeta_m \vert^4         \Big] } d \zeta^{'}_m  d \zeta^{''}_m.~~~\nn\\
\eeqa
 Evaluating the integrals and the product one can get
\be
Z_{negative~modes}^{self-interacting}=   \prod^{{1\over 2  \gamma} -  {1\over 2}  }_{m=0} N_m \Big({ \mu^2 \over g H} \Big)=
N e^{-{(1+\gamma) \over 2\gamma}\ln{g H \over \mu^2} }.
\ee
Taking into account the degeneracy of each negative/unstable mode one can get  
\beqa\label{unrenunsta}
Z_{negative~modes}=\prod^{\CD eg}  Z_{negative~modes}^{self-interacting} &=&  N e^{  -   {\gamma g^2 H^2 \over 4 \pi^2 } {(1+\gamma) \over 2 \gamma}   \ln   { g H  \over  \mu^2 }  V_3 T   } ,
\eeqa
where in the given case (\ref{alpharegeul}) the degeneracy (\ref{degeneracy}) is $\CD eg =  {\gamma g^2 H^2 \over 4 \pi^2 }V_3 T $, and we will get
\be\label{effLagunst1}
\CL^{negative~modes}_{E}=   { g^2 H^2 \over 8 \pi^2 } (1+\gamma)  \ln   { g H  \over  \mu^2 }   .
\ee
The total unrenormalised effective Lagrangian is a sum of the positive modes contribution (\ref{contofnonzeromodes2}) and of the self-interacting negative/unstable modes (\ref{effLagunst1}). 

The above consideration of self-interacting negative/unstable modes shows that the exact integration over these modes can be performed for fully interacting modes.  The general negative/unstable  mode  wave function that includes all degenerate states $\psi_{0m}(n_0,m_0)$  is given  in (\ref{excitedsta}) and (\ref{zeromodeswave}), and the gauge field components (\ref{chargefields}) have the following form:
\beqa\label{unstablemodespar}
&&a_1(\zeta) =i a_2(\zeta)=  \sum^{\CD eg}_{n_0,m_0}\sum^{{1- \gamma \over 2 \gamma}}_{m=0} \zeta_{m}(n_0,m_0) ~(c^+_1-i c^+_2)^{n_0}(d^+_3 -i d^+_0)^{m_0} (d^+_3 +i d^+_0)^{m} \psi_{00}(x)\nn\\
&&a_3(\eta) =i a_0(\eta)=  0,
\eeqa
where $\zeta_{m}(n_0,m_0)$ are the amplitudes  (collective variables) of the negative/unstable modes  that  should be  normalised  similarly to (\ref{generalquartic2}).   The cubic interaction term (\ref{cubic1})  vanishes and  the quartic term gives
\be\label{quarticgen1}
V_E  =    {g^2 \over 2} \int \Big\vert  \sum^{\CD eg}_{n_0,m_0}\sum^{{1- \gamma \over 2 \gamma}}_{m=0}  \zeta_{m}(n_0, m_0) \psi_{0m}(n_0,m_0;x) \Big\vert^4  d^4x,
\ee
and it can be transformed into the following expression: 
\beqa
& V_E  =   \gamma   {g^2  (g H)^2   \over 2} \int  \Big\vert  \sum^{\CD eg}_{n_0,m_0}\sum^{{1- \gamma \over 2 \gamma}}_{m=0}  \zeta_{m}(n_0, m_0) \psi_{0m}(n_0,m_0;y) \Big\vert^4  ~d^4 y  ~, 
\eeqa 
where, as above in (\ref{intzeromodepath}),  we introduced the dimensionless  variables $y_i = x_i  (g H )^{1/2}$, and the field strength dependence completely factorises from the wave function $\psi_{0m}(n_0,m_0;y)$. Now the  integration measure for the negative/unstable modes in the partition function  (\ref{uneqnon}) is
\be
\prod^{negative~modes }_{a \mu} \CD a^{a}_{\mu}  = J \prod^{\CD eg}_{n_0,m_0}   \prod^{ {1 -\gamma\over 2  \gamma} }_{m=0 }  d \zeta^{'}_{m}(n_0, m_0)  d \zeta^{''}_{m}(n_0, m_0) .  
\ee
The  contribution of the negative/unstable modes to the partition function will take the following form: 
\beqa
&Z_{negative~modes}= \\
& \int \exp{\Big\{\sum_{n_0,m_0,m}  g H (1 -\gamma - 2 m \gamma  ) \vert \zeta_m(n_0,m_0)\vert^2    -\gamma    {g^2  (g H)^2  \over 2} \int  \Big\vert  \sum_{n_0,m_0,m}   \zeta_{m}(n_0, m_0) \psi_{0m}(n_0,m_0;y) \Big\vert^4  ~d^4 y  ~\Big\} }~  \nn\\
&J \prod^{\CD eg}_{n_0,m_0}   \prod^{ {1 -\gamma\over 2  \gamma} }_{m=0 }  d \zeta^{'}_{m}(n_0, m_0)  d \zeta^{''}_{m}(n_0, m_0).\nn
\eeqa 
Introducing dimensionless collective variables $ \zeta_{m}(n_0, m_0)  \rightarrow   \zeta_{m}(n_0, m_0)/(gH)^{1/2} $ we will get:
\beqa
&Z_{negative~modes}= \\
& \int \exp{\Big\{\sum_{n_0,m_0,m}  (1 -\gamma - 2 m \gamma  ) \vert \zeta_m(n_0,m_0)\vert^2    -\gamma    {g^2  \over 2} \int  \Big\vert  \sum_{n_0,m_0,m}   \zeta_{m}(n_0, m_0) \psi_{0m}(n_0,m_0;y) \Big\vert^4  ~d^4 y  ~\Big\} }~  \nn\\
&J \prod^{\CD eg}_{n_0,m_0}   \prod^{ {1 -\gamma\over 2  \gamma} }_{m=0 } \Big( { \mu^2 \over g  H  }\Big)  d \zeta^{'}_{m}(n_0, m_0)  d \zeta^{''}_{m}(n_0, m_0).\nn
\eeqa 
The dependence on the field strength completely factorises now from the integral over collective variables  $\zeta_{m}(n_0, m_0)$ and we get
\beqa\label{fullmodecont}
&Z_{negative~modes}=\\
 &N \prod^{\CD eg}_{n_0,m_0}    \prod^{ {1 -\gamma\over 2  \gamma} }_{m=0 }  \Big( {  \mu^2 \over g  H  } \Big)   =
 N  \prod^{\CD eg}_{n_0,m_0}    \Big( {  \mu^2 \over g  H  } \Big) ^{{1 +\gamma\over 2  \gamma} } =N e^{-\CD eg~ {(1+\gamma) \over 2 \gamma}\ln{g H \over \mu^2} }=
N e^{  -   {\gamma g^2 H^2 \over 4 \pi^2 } {(1+\gamma) \over 2 \gamma}   \ln   \Big({ g H  \over  \mu^2 } \Big)   V_3 T },\nn
\eeqa 
 where 
 \beqa
& N= \int \exp{\Big\{\sum_{n_0,m_0,m}  (1 -\gamma - 2 m \gamma  ) \vert \zeta_m(n_0,m_0)\vert^2    -\gamma    {g^2  \over 2} \int  \Big\vert  \sum_{n_0,m_0,m}   \zeta_{m}(n_0, m_0) \psi_{0m}(n_0,m_0;y) \Big\vert^4  ~d^4 y  ~\Big\} } \nn\\
&J \prod^{\CD eg}_{n_0,m_0}   \prod^{ {1 -\gamma\over 2  \gamma} }_{m=0 }    d \zeta^{'}_{m}(n_0, m_0)  d \zeta^{''}_{m}(n_0, m_0).\nn
 \eeqa
Thus the contribution of negative/unstable modes that follows from the expression (\ref{fullmodecont}) is 
\be\label{effLagunst}
\CL_{negative~modes}=   { g^2 H^2 \over 8 \pi^2 } (1+\gamma)    \ln   { g H  \over  \mu^2 },  
\ee
and the contribution of all modes to the  effective Lagrangian  is a sum of (\ref{contofnonzeromodes2}) and (\ref{effLagunst}):
\be\label{efflag}
\CL(\gamma)=  \epsilon^{(1)}_{positive~modes} (\gamma) + \epsilon_{negative~modes}(\gamma).
\ee
The conclusion is that the general chromomagnetic vacuum fields (\ref{alpharegeul}) are also stable because the effective Lagrangian is a real function and is without imaginary terms.  In the case of a  pure chromomagnetic field  $\gamma =0$  the contribution  of positive modes (\ref{contofnonzeromodes2}) can be exactly integrated  (\ref{gammazero}) and together with the contribution of negative/unstable modes (\ref{effLagunst}) will take the following form: 
\be\label{savcond}
\CL^{eff}_E= {  H^2 \over 2 } +{11 g^2 H^2 \over 48 \pi^2} \Big( \ln   { g  H   \over  \mu^2 }   - {1 \over 2} \Big).
\ee
This demonstrates the robustness of the logarithmic structure of the effective Lagrangian  functional \cite{Savvidy:1977as}.  In the case of  (anti)self-dual  field  $\gamma =1$ the sum (\ref{efflag}) reduces to the expression (\ref{euclidlagra}). The conclusion is that the stability of the vacuum fields takes place not only for (anti)self-dual fields but also for the more general covariantly-constant vacuum fields defined in (\ref{alpharegeul}).

\section{\it Consideration in Physical Space-time}

A similar integration of negative/unstable modes can be performed in the Minkowski space-time  \cite{Savvidy:2019grj}, where instead of the Euclidean spectrum (\ref{spectr}) the spectrum of the charge vector bosons in pure chromomagnetic vacuum field $H$ will take the following form \cite{Nielsen:1978rm,Skalozub:1978fy}:
\be\label{magspect}
k^{2}_{0}= k^{2}_{||} + g H (2n+1) \pm 2 gH,
\ee
and it has the  negative/unstable mode 
$
k^2_0 = k_{\vert \vert}^2 - g H  
$
when $k_{\vert \vert}^2 <  g H$. If one ignores for a while the nonlinear interaction of negative/unstable  modes, one can conclude that there is an  imaginary term in  the vacuum energy density \cite{Nielsen:1978rm,Skalozub:1978fy} : 
 \beqa
Im \epsilon^{(1)} = Im ~ {g H \over 4 \pi^2}~ \int^{\infty}_{- \infty} d k_{\vert \vert}  \sqrt{k_{\vert \vert}^2 - g H -i \epsilon} = -  {g^2H^2  \over 8 \pi} . \nn
\eeqa
The conclusion is that  one should take into account the nonlinear interaction of negative/unstable modes. First let us consider the  contribution of the positive modes into the energy density \cite{Nielsen:1978rm,Heisenberg:1935qt}
\be
\epsilon_{positive~modes}= \int^{\infty}_{-\infty} d k_{||}  \Big( \sum^{\infty}_{n=0} \sqrt{k^{2}_{||} +gH (2n+3)} + \sum^{\infty}_{n=1} \sqrt{k^{2}_{||} +gH (2n-1)}   \Big),
\ee
which after the renormalisation will take the following form: 
\be\label{conpos}
\epsilon_{positive~modes}=- {g^2 H^2 \over 48 \pi^2} \Big( \ln   { g H  \over  \mu^2 }   - {1\over 2}\Big) +  {g^2 H^2 \over 8 \pi^2} \Big( \ln   { g H  \over  \mu^2 }   - {1\over 2}\Big).
\ee
Now let us consider the contribution of the negative/unstable  mode by taking into account their nonlinear interaction.    The negative/unstable mode wave function has the following form\footnote{Here the gauge fields are defined as   $a^3_{\mu}$ and $a_{\mu} = (a^1_{\mu} + i a^2_{\mu})/\sqrt{2}$ .} \cite{ Ambjorn:1978ff, Ambjorn:1980ms}:
\be
a_1(x)=i a_2(x) = \int {d k_2 \over 2 \pi } 
e^{- {1\over 2} g H (x_1 - k_2/g H)^2 + i k_2  x_2 }  {\phi_{k_{2}}(x_3,x_0) \over 2^{1/4}} ,
\ee
where $\phi_{k_{2}}(x_3,x_0)$ is the dimensionless amplitude (collective variable)  of negative/unstable modes,  and it is analogous to the amplitudes $\zeta_{m}(n_0,m_0)$ in the Euclidean path integral formulation  (\ref{unstablemodespar}).  The energy spectrum (\ref{magspect}) does not depend on the continuous momentum variable $k_2$ and exposes  the true infinite  degeneracy (\ref{degeneracy}) of the spectrum  (\ref{magspect}).   If one introduces the dimensionless variables $k_{\mu} \rightarrow k_{\mu}/ \sqrt{gH}$, $x_{\mu} \rightarrow x_{\mu}  \sqrt{gH}$  in the wave function, then it takes the following form \cite{Savvidy:2019grj}:  
\be\label{unstablemode}
a_1(x)=i a_2(x)  =    ( g H )^{1/2}   \int {d k_2 \over 2 \pi} e^{-{1\over 2}(x_1 - k_2 )^2 + i k_2 x_2}  {\phi_{k_{2}}(x_3,x_0) \over 2^{1/4}}.
\ee
In terms of the dimensionless variables the part of the action that corresponds to the negative/unstable mode \cite{ Ambjorn:1978ff, Ambjorn:1980ms}  will take the following form \cite{Savvidy:2019grj}: 
\be\label{unstablelagrangian}
S_{negative\ mode}  =  \int {d k_{2} \over \sqrt{2 \pi }} d x_0 d x_3 \Big(  \vert \partial_{\mu} \phi_{k_{2}} \vert^2 +  \vert  \phi_{k_{2}} \vert^2 
- {1\over 2} g^2  \int {d p d q \over (2 \pi)^2 }  e^{-{p^2 +q^2\over 2}} \phi^*_{k_2 +p} \phi^*_{k_2 +q} \phi_{k_2} \phi_{k_2 +p+q}\Big),
\ee
and it contains the tachyonic  mass term  $\vert  \phi_{k_{2}} \vert^2$ and the positive definite  quartic  interaction term that provides a convergence of the path integral over the negative/unstable  modes. In terms of the dimensionless variable  the dependence on the chromomagnetic vacuum field completely factorises  from the action (\ref{unstablelagrangian}) while it appears only in front of the wave function  (\ref{unstablemode}) and in the path integral measure   $\prod^{negative\ modes}_{k_2} ({\mu^2 \over gH})^{1/2} \CD  \phi_{k_{2}} $. Thus  the contribution of the negative/unstable mode into the partition function  appears only through the integration measure and the degeneracy of the negative/unstable modes: 
\be\label{negmod}
Z_{negative~mode}   =  N \Big({ \mu^2  \over g H } \Big)^{{i\over 2}  ({gH \over 2\pi})^2 V_3 T} = N e^{- i {g^2 H^2 \over 8 \pi^2}  \log {g H \over \mu^2} V_3 T ~}.
\ee 
The vacuum energy density is a real function of the chromomagnetic vacuum field and is a sum of positive (\ref {conpos})  and negative/unstable mode (\ref{negmod}) contributions: 
\be\label{negamodes}
\epsilon= {H^2\over 2}+ \epsilon^{(1)}_{negative~mode}+\epsilon_{negative~mode}=  {H^2\over 2} +{g^2 H^2 \over 48 \pi^2} \Big( \ln   { g H  \over  \mu^2 }   - {1\over 2}\Big).
\ee
This confirms that the energy density does not have an imaginary part  \cite{Savvidy:1977as, Savvidy:2019grj}. 

One can also include the interaction of the negative/unstable mode with the neutral mode $a^3_{\mu}$ of the Yang-Mills field. The interaction term was derived in \cite{ Ambjorn:1978ff, Ambjorn:1980ms}, and in terms of dimensionless variables introduced above it will take the following form:
\beqa
&S^{neutral~mode}_{negative~mode} = \int d x_3 d x_0 \int {d k_1 d k_2 \over (2 \pi )^{3/2}} e^{-(k^2_1 +k^2_2)/4} \int {d k^{'}_2 d k^{''}_2 \over 2 \pi } e^{-k^2_1 -k^2_2} \delta(k_2 +k^{'}_2 - k^{''}_2 )e^{i k_1(k^{'}_2 + k^{''}_2)/2}\nn\\
&\Big[-ig a^3_{\mu}  \phi^*_{k^{''}_2}  \partial_{\mu} \phi_{k^{'}_2 }  -g^2 (a^3_{\mu})^2   \phi^*_{k^{''}_2}   \phi_{k^{'}_2 }    \Big].
\eeqa
This part of the Yang-Mills action also does not depend on the field strength. The contribution of the unstable and neutral  modes into the partition function  appears only through the integration measure and the degeneracy of these states,  and therefore does not alter the previous result (\ref{negamodes}).

\section{\it Chromomagnetic Gluon Condensate } 

We showed that the effective Lagrangian of the SU(N) Yang-Mills  theory does not have an imaginary term and has the following form \cite{Savvidy:1977as}:
\be\label{YMeffective}
\CL  =  
-\CF - {11  N \over 96 \pi^2}  g^2 \CF \Big( \ln {2 g^2 \CF \over \mu^4}- 1\Big),~~~~~~~~~~~\CF= {1\over 4} G^a_{\mu\nu}G^a_{\mu\nu} >0 . 
\ee
It follows from this expression  that the chromomagnetic magnetic induction $\vec{\CB}_a$  of the YM vacuum  is
\be\label{palarYM}
\vec{\CB}_a = - {\partial \CL \over \partial \vec{\CH}_a} 
= \vec{\CH}_a \Big[1+ {11 g^2  N \over 96 \pi^2}\log{g^2 \vec{\CH}^2_a \over \mu^4} \Big]~~~= \mu_{vac}~   \vec{\CH}_a 
\ee
and that the YM vacuum behaves as a paramagnet with a magnetic permeability of the following form \cite{PhDTheses}:
\be\label{permeabilityYM}
\mu_{vac} = -{\partial \CL \over \partial \CF}  =1+ {11 g^2  N \over 96 \pi^2}\log{g^2 \vec{\CH}^2_a \over \mu^4} ~ = {11 g^2  N \over 96 \pi^2}\log{g^2 \vec{\CH}^2_a \over \Lambda^4_{S}} .
\ee
The paramagnetism of the YM vacuum at  $g^2 \vec{\CH}^2_a \geq \Lambda^4_{S}$ means that there is an amplification of the chromomagnetic vacuum fields very similar to the Pauli paramagnetism, an effect associated with the polarisation of the electron spins. In YM theory the polarisation of the virtual vector boson spins is responsible for the vacuum fields amplification. This also can be seen from the vacuum energy density  with its new minimum outside of the perturbative vacuum $ < \CF > =0$ at the renormalisation 
group invariant field strength \cite{Savvidy:1977as}
\be\label{chomomagneticcondensate}
 \langle 2  g^2 \CF \rangle_{vac}=    \mu^4  \exp{(-{32 \pi^2 \over 11 g^2(\mu) })} = \Lambda^4_{S}
\ee
or, in terms of the strong coupling constant,
\be
\langle {\alpha_s \over \pi} G^2_{\mu\nu} \rangle_{vac} = \langle {g^2 \over 4 \pi^2} G^2_{\mu\nu} \rangle_{vac}= 
{\Lambda^4_{S} \over 2 \pi^2}~. 
 \ee
Using the relation $\Lambda_{P} /\Lambda_{S} \approx 2.11$ derived in \cite{Nielsen:1978zg} and that $\Lambda_{P} \approx 200 MeV$ one can get  \cite{ Savvidy:2019grj}
$$
\langle {\alpha_s \over \pi} G^2_{\mu\nu} \rangle_{vac} \approx ~0.0000041~ GeV^4.
$$ 
As we have seen, a large class of chromomagnetic vacuum fields is stable and indicates that the Yang-Mills vacuum is a highly degenerate  quantum state. It is also appealing that even a larger class of alternative vacuum fields have also been considered in the recent publications \cite{Baseyan, Nielsen:1979vb, Natalia, SavvidyKsystem, SavvidyKsystem1, SavvidyPlanes, Banks, Anous, Cho:1979nv, Kim:2016xdn, Milshtein:1983th, Olesen:1981zp,  Apenko:1982tj, Savvidy:2020mco}, and some of them expose a natural chaotic behaviour.  
In this respect one should ask whether there exist physical systems that have high degeneracy of the vacuum state. Turning to the statistical spin systems, one can observe that the classical 3D Ising system has a double degeneracy of all its excited states and of the vacuum state. It is this symmetry that allows to construct a dual gauge invariant representation  of the 3D Ising model \cite{Wegner:1971app}. The extensions of the 3D Ising model that have a direct ferromagnetic  and  one quarter  of the next  to nearest  neighbour  antiferromagnetic interaction constructed in \cite{Savvidy:1993ej}, as well as a model with a zero intersection coupling constant ($k=0)$ \cite{Savvidy:1993sr, Savvidy:1994sc,Pietig:1996xj, Pietig:1997va}, have {\it exponential degeneracy of the vacuum state}.  In recent publications this symmetry was referred as the subsystem symmetry \cite{Vijay:2016phm}. This higher symmetry allows to construct the dual representations of the same systems  and in various  dimensions \cite{Savvidy:1993sr, Savvidy:1994sc,Savvidy:1994tf}.  As a  consequence of the high degeneracy of the vacuum state,  these systems have rich physical properties, including a glass behaviour \cite{Lipowski, Sherrington}  and exotic fracton excitations \cite{Vijay:2016phm}.

\section{\it Large N Behaviour }

Let us consider the  behaviour of the effective Lagrangian from the renormalisation group point of view and in the limit of large $N$  \cite{tHooft:1973alw}.  When  $\CG = \vec{\CE_a} \vec{\CH_a} =0 $, we have  
\be\label{effectivecoupling}
\CM(t,g) = {\partial \CL \over \partial \CF}= - {g^2 \over \bar{g}^2(t)},~~~~~~{d \bar{g} \over dt } = \bar{\beta}(\bar{g}) .
\ee
The   vacuum magnetic permeability  (\ref{palarYM}) will take the following form  \cite{PhDTheses}:
\be\label{permeabilityfulYMl}
{\it \mu}_{vac}~ = ~ {g^2 \over \bar{g}^2(t)}  ,~~~~~~~~~\CG=0. 
\ee
The Callan-Symanzik beta function can be calculated by using (\ref{YMeffective}): 
\be\label{betafunction}
\bar{\beta}=  {1\over 2}g {\partial \CM \over \partial t} \vert_{t =0}= -{11  N \over 96 \pi^2}g^3,
\ee
and the effective coupling constant as a function of the vacuum  field strength has the following form:
\be\label{effectivecouplingYM}
\bar{g}^2(\CF) = {g^2  \over 1+ {  11 g^2  N \over 96 \pi^2} 
  \ln{2 g^2 \CF \over \mu^4}  }. 
\ee
 Let us consider the field strength  $\CF_{0}$ at which the vacuum energy density  vanishes  $\epsilon(\CF_0)=0$:
\be\label{intersectionspoint}
2 g^2 \CF_{0}=   \mu^4 \exp{(1-{96 \pi^2 \over 11 g^2 N} )} = e  \langle 2  g^2 \CF \rangle_{vac}.
\ee 
The effective coupling constant (\ref{effectivecouplingYM}) at this field strength has the value 
\be\label{intersection}
\bar{g}^2(\CF_0) = {96 \pi^2   \over   11 N}~. 
\ee
It follows  that the effective coupling constant at the intersection point $\CF_0$ is small: \be\label{effcoupling}
\bar{g}^2(\CF_0) = {96 \pi^2   \over   11 N}   \ll 1 ~~~\text{if}~~~ N \gg   {96 \pi^2   \over   11} ~.
\ee
The energy density curve $\epsilon(\CF)$   intersects the zero energy density level  at a nonzero angle  $\theta$:
\be\label{intersectionangle}
\tan\theta={ 11 g^2 N \over  96 \pi^2 }  ~>~ 0.
\ee
This means that  i) the vacuum state is  below the perturbative vacuum  and that ii) there is a nonzero vacuum field condensate.  Now, the question is, how far into the infrared region one can continue the energy density curve by using the perturbative results?  Let us consider the vacuum fields that are close to the infrared pole. This can be achieved  by using the following parametrisation: 
\be\label{infraredpole}
  \CF_{\alpha}=e^{1-\alpha} \langle  \CF \rangle_{vac},
\ee
where the parameter $\alpha \leq 1$.  When $\alpha$ is close to one, we will have  $ \CF_{\alpha} ~\rightarrow  ~ \langle  \CF \rangle_{vac}$.  At this field the value of the effective coupling constant (\ref{effectivecouplingYM})  tends to zero,
\be\label{infraredpole1}
\bar{g}^2(\CF_\alpha) = {96 \pi^2   \over   11 N (1-\alpha)} \rightarrow  0,
\ee
if the product $N (1-\alpha) $  is large and the t'Hooft coupling constant $g^2 N = \lambda$ is fixed and is small \cite{tHooft:1973alw}.
 The energy density curve can be continuously  extended infinitesimally close to the value of the vacuum field $\langle  \CF \rangle_{vac}$ in this limit.    

One can  analyse the effective coupling constant  (\ref{intersection})  and the intersection point (\ref{intersectionspoint}) at the two-loop level. The two-loop\footnote{The beta   function (\ref{effectivecoupling}) coefficients  $\bar{\beta} = -\beta_1 g^3 - \beta_2 g^5 +..$  are  given by $\beta_1 = {11 N   \over 6 (4\pi)^2 } $ and $\beta_2 = {34  N^2 \over 6 (4 \pi)^4 } $ \cite{Jones:1974mm,Caswell:1974gg}. } effective Lagrangian has the following form \cite{PhDTheses}:
\be
\CL  =  
-\CF - \Big({11   \over 6 (4\pi)^2 } g^2 N + {34   \over 6 (4 \pi)^4 } (g^2 N)^2 \Big)  \CF \Big( \ln {2 g^2 \CF \over \mu^4}- 1\Big).   
\ee
The field at the intersection point (\ref{intersectionspoint}) is shifted  by an exponentially small correction
\be\label{effectivecouplingYM1}
2 g^2 \CF_{0}= \mu^4  \exp{\Big(1-{96 \pi^2 \over 11 \lambda} \cdot{1\over 1 + {17 \over 88 \pi^2} \lambda } \Big)}.
\ee
At this field the effective coupling constant is smaller by the factor $1/  1 + {17 \over 88 \pi^2} \lambda  $:
\be\label{intersection1}
\bar{g}^2(\CF_0)  ={96 \pi^2   \over   11 N} \cdot {1   \over   1 + {17 \over 88 \pi^2} \lambda  } \ll 1,
\ee
and  the inequality (\ref{intersection1}) is fulfilled at smaller values of $N$ than in the first approximation (\ref{effcoupling}).   The  chromomagnetic  condensate in the two-loop approximation will take the following form:   
\be\label{twoloopcond}
\langle 2  g^2 \CF \rangle_{vac}=  
 \mu^4  \exp{\Big(~  -{1\over  \beta_1 g^2} \Big[1 - {\beta_2 g^2 \over \beta_1} \ln(1 +   {\beta_1 \over \beta_2 g^2} )\Big]  ~       \Big)}.
\ee
The high-loop corrections can be obtained by using the renormalisation group results (\ref{effectivecouplingel}), (\ref{effectivecoupling}). Then the value of the chromomagnetic condensate is \cite{Savvidy:1977as}:
\be\label{chomomagneticcondensaterenormgroup}
\langle 2  g^2 \CF \rangle_{vac}=    \mu^4  \exp{\Big(2 \int^{\infty}_{g}{d g \over \bar{\beta}(g)}\Big)}.
\ee
 The expression (\ref{twoloopcond}) is  recovered at the two-loop level.

\section{\it Conclusion }

It is remarkable that the zero mode contribution that was calculated in terms of  infrared regularisation of the spectrum (\ref{infracon}),  by integration over the self-interacting zero modes in (\ref{zeromodepro1}) and then by integration over fully interacting zero modes in (\ref{fullzeromodecont}), all lead to the same result indicating the robustness of the logarithmic structure of the effective Lagrangian  and that it is without an imaginary part \cite{Savvidy:1977as}. A similar phenomenon took place when we were considering the contribution of the negative/unstable modes by integrating over self-interacting (\ref{unrenunsta}) and fully interacting modes (\ref{fullmodecont}).  It is interesting to investigate  to what extent this behavioural robustness is rooted in the entropy factor,  through the degeneracy of the quantum states  and through the scaling invariance of the Yang-Mills action.

One can consider the above approach of calculating the effective action as an alternative loop expansion of the effective action. The expansion is organised by rearranging the perturbative expansion in a vacuum field so that the interaction of all negative/unstable modes is included into the propagator of the gauge field $G(x,y;A) $ and  the  loop expansion is performed in terms of the  cubic and quartic cross-mode  vertices between positive/stable and negative/unstable modes  of the YM action \cite{Savvidy:2019grj}.  Some technical details are presented in the Appendixes.

In conclusion I would like to thank R. Kirschner,  J. Zahn and M. Bordag for discussions of the YM vacuum stability and the kind hospitality at the Institute of Theoretical Physics of the Leipzig University where this work was completed.  This work was supported by the Alexander von Humboldt Foundation grant GRC 1024638 HFST. I would like to thank J. Ambjorn, H. B. Nielsen and K.Savvidy for many stimulating discussions of the structure of the QCD vacuum.

\section{\it Appendix A.  Degeneracy of Eigenstates }
A component  representation of the operator $H_0=-\nabla_{\mu} \nabla_{\mu}$ is given in (\ref{theh0}).
 By using the operators (\ref{newoperators}) one can find that  
\beqa
&c^+_1 + ic^+_2  =  -\partial_1- i \partial_2 + {g\over 2} H (x_1 +i x_2), ~~~~
 c_1 - ic_2  =   \partial_1- i \partial_2 + {g\over 2} H (x_1 -i x_2),\nn\\
&[c^+_1 + ic^+_2~; ~c_1 - ic_2]= -2g H 
\eeqa
and that the magnetic part of the $H_0$ is
\beqa
&&(c^+_1 + ic^+_2)(c_1 - ic_2)=  -\partial^2_1- \partial^2_2 + ig H (  x_2 \partial_1 -  x_1 \partial_2) + {g^2 \over 4} H^2 ( x^2_2+  x^2_1) -g H. 
\eeqa
The vacuum wave function is defined as
\be
(c_1 - ic_2)\psi_0 = 0,~~~~~~\psi_0=  e^{- {g H\over 4}(x_1^2 + x_2^2 )   }  
\ee
and the wave functions of the excited states (\ref{excitedsta}) are
\be
\psi_n = (c^+_1 + ic^+_2)^n \psi_0 = (g H)^n(x_1+i x_2)^n \psi_0.
\ee
The set of operators (corresponding to the centre of the cyclotron motion of charge particle in a magnetic field) that defines the degeneracy of the eigenstates is
\beqa
&  c^+_1 -i c^+_2=-\partial_1+ i \partial_2 + {g\over 2} H (x_1 -i x_2),~~~~
   c_1+i c_2 =   \partial_1+ i \partial_2 + {g\over 2} H (x_1 +i x_2),\nn\\
&[c^+_1 - ic^+_2~; ~c_1 + ic_2]= -2g H.
\eeqa
The wave functions of the degenerate vacuum state are
\beqa\label{degeneratewave}
~~~\psi( n_0) = (c^+_1 -i c^+_2)^{n_0}  \psi_{0}= (g H)^{n_0} (x_1- ix_2)^{n_{0}}  \psi_{0}.
\eeqa
They are the eigenstates of the operator $L =i (x_1 \partial_2 -x_2 \partial_1)$:
\be
L~\psi( n_0) = n_0~ \psi( n_0).
\ee
All of them have the ground state eigenvalue in (\ref{H0spect}): 
\be
\lambda_0 = g H . 
\ee
The normalisation of the wave functions $\psi_{n_0}$ is
\beqa\label{normalisa}
C_{n_0}&=&\int d x_1 d x_2 \vert \psi(n_0) \vert^2=  2^{n_0+1} \pi n_0! (g H)^{n_0-1}, 
\eeqa
and these wave functions (\ref{degeneratewave}) are orthogonal:
\be
  \int d x_1 d x_2   \psi^*(n_0) \psi(\bar{n}_0)  = C_{n_0} \delta_{n_0,\bar{n}_0} .
\ee
The average size of the orbit of the degenerate state $\psi(n_0)$  is
\be
\langle r^2 \rangle=\langle  x_1^2 +  x_2^2 \rangle={1\over C_{n_0}}\int d x_1 d x_2  (  x_1^2 +  x_2^2) \vert \psi(n_0) \vert^2 = {2(n_0+1) \over g H}.  \nn
\ee
It follows then that the degeneracy, the number of the charge particle orbits of the size less than $r^2 \leq R^2$, is
\be
n_0+1 \leq {g H \over 2} R^2 =  {g H \over 2 \pi }  \pi R^2 = {g H \over 2 \pi } A_{12},
\ee
 where $A_{12}$ is two-dimensional area on the plane $(x_1,x_2)$. The same is true for the chomoelectric part  of the $H_0$ operator (\ref{theh0}) on the plane $(x_3,x_0)$:
\be
m_0+1 \leq {g E \over 2} R^2 =  {g E \over 2 \pi }  \pi R^2 = {g E \over 2 \pi } A_{03},
\ee
thus the total degeneracy is 
\be\label{degeneracy}
\CD eg=\Big({g H \over 2 \pi } \Big) \Big({g E \over 2 \pi } \Big)V_3 T,
\ee
where $V_3$ is a three-volume and $T$ play the role of a time-like parameter in (\ref{euclid0}).

\section{\it Appendix B. Self-Interaction of Zero Modes}
The zero modes are the solutions of the equation (\ref{eigenva}) 
\be
H_0  a_{\nu} + 2 i g  F_{\mu\nu} a_{\mu}=0
\ee
with $\lambda=0$, and the solution (\ref{zero}) corresponding to the lowest state with $n_0=m_0=0$ is:
\beqa\label{zeroditections1}
&a^1_1 = \xi^{'} \psi_{00},~~&a^2_1 = \xi^{''} \psi_{00},~~~~a^3_1=0\nn\\
&a^1_2 = \xi^{''} \psi_{00},~~&a^2_2 = -\xi^{'} \psi_{00},~~a^3_2=0\nn\\
&a^1_3 = \eta^{'} \psi_{00},~~&a^2_3 = \eta^{''} \psi_{00},~~~~a^3_3=0\nn\\
&a^1_0 = \eta^{''} \psi_{00},~~&a^2_0 = -\eta^{'} \psi_{00},~~a^3_0=0,
\eeqa
where $\xi_{00} = \xi^{'}  +i  \xi^{''}  $ and $\eta_{00} = \eta^{'}  +i  \eta^{''}  $.The interaction cubic term (\ref{nonlinear}) vanishes on this solution:
\be\label{cubic}
V^{(3)}_E=- g \epsilon^{abc} a^{b}_{\nu}a^{c}_{\mu} \nabla^{ad}_{\mu} a^{d}_{\nu}=0,
\ee
and the quartic term will take the following form:
\be\label{quartic}
V^4_E= {g^2 \over 4} \int d^4x (\epsilon^{abc}a^{b}_{\mu}a^{c}_{\nu})^2=  {g^2 \over 4}\int d^4x ((a^{a}_{\mu} a^{a}_{\mu})^2
-(a^{a}_{\mu}a^{a}_{\nu})^2 )
=  {g^2 \over 2}(\xi^{' 2}+\xi^{'' 2}+\eta^{' 2}+\eta^{'' 2})^2 \int d^4x  \vert \psi_{00} \vert^4,
\ee
where 
$
 (a^{a}_{\mu} a^{a}_{\mu})^2=
4(\xi^{' 2}+\xi^{'' 2}+\eta^{' 2}+\eta^{'' 2})^2,~
(a^{a}_{\mu}a^{a}_{\nu})^2 = 2(\xi^{' 2}+\xi^{'' 2}+\eta^{' 2}+\eta^{'' 2})^2.
$
We have to calculate the quartic integral for the lowest state  $\psi_{00}$:
\beqa\label{normalisa1}
&&{1\over C^4_0}\int d x_1 d x_2  d x_3 d x_0  \vert \psi_{00} \vert^4= \Big({g H  \over 2 \pi}\Big)^4  \Big({\pi \over g H}\Big)^2 = \Big({g H  \over 4 \pi}\Big)^2 .  
\eeqa
The quartic term for the general zero mode wave function (\ref{zeromodeswave}), (\ref{zero}) will take the following form:
\be\label{quarticgen}
V^4_E =  {g^2 \over 2}(\vert \xi_{n_0 m_0} \vert^2 +\vert \eta_{n_0 m_0} \vert^2)^2 \int d^4x \vert \psi_{n_0 m_0} \vert^4,
\ee
and then  using the integral  of the wave function (\ref{degeneratewave}) one can get:
\be
  \int d x_1 d x_2 \vert \psi(n_0) \vert^4    =    \pi \Gamma(2n_0+1) (g H)^{2n_0-1} .
\ee
The zero mode (\ref{zero})   self-interaction term  (\ref{nonlinear})  has the following form:
\be\label{quarticgen}
V^{zero~modes}_E =  {g^2 \over 4}  \int d^4x   (\epsilon^{abc}a^{b}_{\mu}a^{c}_{\nu})^2 = {g^2 \over 2}(\vert \xi_{n_0 m_0} \vert^2 +\vert \eta_{n_0 m_0} \vert^2)^2 ~\int d^4x \vert \psi_{00}(n_0, m_0;x) \vert^4,
\ee
where $\psi_{00}(n_0, m_0;x) =(g H)^{n_0} (x_1- ix_2)^{n_{0}} (g E)^{m_0} (x_3- ix_0)^{m_{0}}  \psi_{00}  $.
For the normalised wave function  one can  get: 
\be\label{generalquartic2}
{1\over C^2_{n_0} } {1\over C^2_{m_0} } \int d^4 x   \vert \psi_{00}(n_0, m_0;x) \vert^4  = \Big({g H  \over 4 \pi}\Big)^2 {\Gamma(n_0 +1/2) \Gamma(m_0 +1/2) \over \pi \Gamma(n_0+1) \Gamma(m_0+1)}.
\ee

\section{\it Appendix C. Sum of Eigenvalues}

The spectral sums (\ref{contofnonzeromodes}) and (\ref{contofnonzeromodes2}) can be evaluated by using the following formulas: 
\beqa\label{sum1}
&&\sum^{\infty}_{n,m=0} e^{-  g H (2 n +2 \gamma m +3+\gamma) s} =  e^{-2 g H  s } {1\over 4 \sinh(g H s)} {1\over \sinh(g  \gamma H s)}\nn\\
&&\sum^{\infty}_{n,m=0}e^{- g H (2 n +2\gamma m +1+3 \gamma) s} =  e^{-2 g \gamma H  s } {1\over 4 \sinh(g H s)} {1\over \sinh(g  \gamma H s)}\nn\\
&& \sum^{\infty}_{n,m=0}e^{-  g H (2n +2\gamma m +1- \gamma) s} =  e^{2 g \gamma H  s } {1\over 4 \sinh(g H s)} {1\over \sinh(g  \gamma H s)}\nn\\
&&\sum^{\infty}_{n,m=0}  e^{- g H (2 n +2 \gamma m -1 +\gamma) s} =  e^{2 g \gamma H  s } {1\over 4 \sinh(g H s)} {1\over \sinh(g   \gamma H s)}\nn\\
&&  \sum^{\infty}_{n,m=0} e^{- g H (2 n +2 \gamma m +1+\gamma) s} =   {1\over 4 \sinh(g H s)} {1\over \sinh(g  \gamma  H s)}\nn\\
&& \sum^{{1\over 2 \gamma} -  {1\over 2}  } _{m=0}e^{- g H ( 2 \gamma m -1 +\gamma) s}=
e^{g H s(1-\gamma)} {1- e^{-g H s(1+\gamma)}  \over 1- e^{-2 g \gamma H s  }}.
\eeqa
The integrals appearing in the effective Lagrangian   have the following form:
\beqa\label{integral}
&&\int^{\infty}_{0}  
  {   ds \over  s^{1-k} ~ \sinh^2( a s) }= {4  \over (2 a)^k}  \Gamma(k)  \zeta(k-1), \\ 
&& \int^{\infty}_{0}  
 {    \cosh(b s) ds \over  s^{1-k}~ \sinh( a s) }= {\Gamma(k) \over (2 a)^k} \Big[ \zeta(k, {1\over 2}(1- {b \over a}) + \zeta(k, {1\over 2}(1+ {b \over a})\Big], ~~~b \neq a,\nn\\
&&\int^{\infty}_{0}  
  {   ds \over  s^{1-k} ~ \sinh( a s) }= {2^k -1  \over 2^{k-1} a^k}  \Gamma(k)  \zeta(k), ~~~~
 \int^{\infty}_{0}  
 {    \cosh(a s) ds \over  s^{1-k}~ \sinh^2( a s) }= {2^{k-1} -1  \over 2^{k-2} a^k}  \Gamma(k)  \zeta(k-1) \nn\\ 
&& \int^{\infty}_{0}  
 {    \sin(a s) ds \over  s^{1-k}}=  {\Gamma(k) \over a^k} \sin{k \pi \over 2},~~~~~~~~~~~~~~
\int^{\infty}_{0}  
 {    \cos(a s) ds \over  s^{1-k}}= {\Gamma(k) \over a^k} \cos{k \pi \over 2} ,\nn
\eeqa
where  $k$ can be considered as a dimensional regularisation parameter  and the integrals should be calculated in the limit $k \rightarrow 0$ \cite{PhDTheses}.  

\section{\it Appendix E. Renormalisation Group  }
The exact expression for the effective Lagrangian can be derived by using the renormalisation group equation \cite{Savvidy:1977as,Matinyan:1976mp}. The effective action $\Gamma$ is a renormalisation group invariant quantity:  
\beqa
 \Gamma &=& \sum_n \int dx_1...dx_n \Gamma^{(n) a_1...a_n}_{~~~\mu_1...\mu_n}(x_1,...,x_n) A^{a_1}_{\mu_1}(x_1)...
 A^{a_n}_{\mu_n}(x_n),  \nn
 \eeqa 
because the vertex functions and gauge fields  transform as follows: 
\beqa
\Gamma^{(n)~ a_1...a_n}_{r~~~\mu_1...\mu_n}  = Z^{n/2}_{3}\Gamma^{(n)~~a_1...a_n}_{un~~~\mu_1...\mu_n},~~~~~
A^{a}_{\mu}(x)_r = Z^{-1}_{3}A^{a}_{\mu}(x)_{un},~~~~
g_r = Z^{1/2}_{3} g_{un} .~~~\nn
\eeqa
The renormalisation group equation takes the form 
\be
\{ \mu^2 {\partial \over \partial \mu^2} + \beta(g) {\partial \over \partial g} + \gamma(g) 
\int d^4x A^{a}_{\mu}(x) {\delta \over \delta A^{a}_{\mu}(x)} \} \Gamma =0, \nn
\ee
where $\beta(g)$ is the Callan-Symanzik beta function, $\gamma(g)$ is the anomalous dimension.   When  $\CG = \vec{\CE}_a \vec{\CH}_a =0 $ it reduces to the form
\be
\{ \mu^2 {\partial \over \partial \mu^2} + \beta(g) {\partial \over \partial g} + 2 \gamma(g) 
\CF {\partial \over \partial \CF}  \}  \CL =0,\nn
\ee
where in the covariant background  gauge $\beta = - g \gamma$ \cite{PhDTheses}. By introducing a dimensionless quantity  
\be\label{derivative}
\CM(g,t) = {\partial \CL \over \partial \CF},~~~~~~~~~t = {1\over 2}\ln(2 g^2 \CF/ \mu^4),  
\ee
one can get 
\be\label{renormequation}
\{ - {\partial \over \partial t} + \bar{\beta}(g) {\partial \over \partial g} + 2 \bar{\gamma}(g)  \}  \CM(g,t) =0, 
\ee
where 
\be\label{fullbeta}
\bar{\gamma} =  {\gamma \over 1- \gamma} ,~~~~~~~\bar{\beta} =  {\beta \over 1- \gamma}
\ee
and (\ref{normalosation1}) plays the role of the boundary condition:
\be\label{boundarycondition}
\CM(g,0) = -1.
\ee
From equations (\ref{renormequation}) and (\ref{boundarycondition}) it follows that 
\be\label{beta}
\bar{\gamma}=  -{1\over 2}{\partial \CM(g,t) \over \partial t} \vert_{t=0}, ~~~~
\bar{\beta}=   {1\over 2} g {\partial \CM(g,t) \over \partial t} \vert_{t=0}~.
\ee
The solution of the renormalisation group equation (\ref{renormequation}) in terms of the effective coupling constant $\bar{g}(g,t)$ with the boundary condition $\bar{g}(g,0) =g$  has the following form \cite{Savvidy:1977as,Matinyan:1976mp}:
\be\label{effectivecouplingel}
  {\partial \CL \over \partial \CF} = - {g^2 \over \bar{g}^2(t)},~~~~~~~~~{d \bar{g} \over dt } = \bar{\beta}(\bar{g}) .
\ee
The behaviour of the effective Lagrangian at large fields is similar to the behaviour of the gauge theory at a large momentum. 
It follows that $\CM(g,t)$ is completely determined for all values of $t$ in terms of its first derivative (\ref{beta}) at $t=0$.  The above results allow to obtain the renormalisation group expressions  for the physical quantities 
considered in a one-loop approximation. With these expressions in hand one can calculate different observables of physical interest that will include  the vacuum energy density and pressure, the magnetic permeability, the  effective coupling constants, and their behaviour as a function of vacuum  fields \cite{Savvidy:2021ahq}.

 \section{\it Appendix D. Euclidean Formulation of Quantum Mechanics }

The matrix elements of the evolution operator $e^{i H t}$ are defined in terms of path integral over trajectories in the physical space-time as \cite{Feynman:1948ur}: 
\beqa\label{largeT1}
&&\langle \vec{x}^{~'} \vert e^{i H T} \vert \vec{x} \rangle = \sum_n e^{-i E_n T} \psi_n(\vec{x}^{'}) \psi^{*}_n(\vec{x})  = N \int^{ \vec{x}^{~'}}_{\vec{x}} \CD \vec{x}(t) e^{i S[\vec{x}(t)]}. 
\eeqa
The integration  runs over all trajectories $\vec{x}(t)$ that start at  time $t =0$ in the point $\vec{x}$ and end at  time $t=T $ in the point $ \vec{x}^{'}$. The path integral is defined in terms of a sum over the physical space-time trajectories. The vectors $\psi(\vec{x}) =\langle x \vert n \rangle$   are the Schr\"odinger wave functions. The $H$ is the Hamiltonian of the system  and $\psi_n(\vec{x})$
is the eigenfunction  $H \psi_n =E_n \psi_n$. 

With the rotation $t \rightarrow -i \tau $ to the Euclidean  time the matrix elements transform to the matrix elements of the operator $e^{-H T}$  and are represented in terms of Euclidean path integral \cite{Coleman}:
\beqa\label{euclid}
\langle \vec{x}^{'} \vert e^{- H T} \vert \vec{x} \rangle = 
\sum_n e^{- E_n T} \psi_n(\vec{x}^{'}) \psi^{*}_n(\vec{x})=    N \int^{ \vec{x}^{~'}}_{\vec{x}} \CD\vec{x}(\tau) e^{-S_E[x(\tau)]}. 
\eeqa
The path integral is defined as an integral in unphysical Euclidean space. The integration runs over all Euclidean trajectories $\vec{x}(\tau)$ that start at Euclidean time $\tau = 0$ in the point $\vec{x}$ and end at Euclidean time $\tau = T$ in the point $ \vec{x}^{'}$. The $S_{E}$ is the Euclidean action associated with a given trajectory:
\be
S_{E}= \int^{T}_{0} d\tau \Big( {1\over 2} m \dot{\vec{x}}^2  + V(\vec{x})\Big).
\ee
{\it The boundary conditions in both of the formulations are identical}, the difference is in the geometry of the trajectories.  In the first case the trajectories are in a real physical space-time where a particle moves in a potential $V$ bounded from below.  In the  Euclidean formulation a particle "moves"  in a potential $-V$ which is unbounded from below,  its "trajectories" are in unphysical Euclidean space and do not directly correspond to the trajectories in a physical space.   At large values of $T$  the leading term in (\ref{euclid})  defines the ground state energy $E_0$ and the corresponding eigenfunction $\psi_0(\vec{x})$:
\beqa\label{euclid1} 
\sum_n e^{- E_n T} \psi_n(\vec{x}^{'}) \psi^{*}_n(\vec{x}) \rightarrow  e^{- E_0 T} \psi_0(\vec{x}^{'}) \psi^{*}_0(\vec{x}) . 
\eeqa 
The above relation  leads to the following observation: In some cases the Euclidean path integral can readily be evaluated in the semiclassical limit when it is dominated by the stationary "trajectory" of $S_E$.   Considering  a standard oscillator with boundary condition $x^{'}=x=0$  one can find that the  stationary "trajectory"  is simply  $\hat{\vec{x}} =0$ and leads to the relation \cite{Coleman}
\be\label{groundstate}
e^{- E_0 T} \vert \psi_0(0) \vert^2 = \Big({\omega \over \pi}\Big)^{1/2} e^{-\omega T /2}
\ee
allowing to extract the value of the ground state energy $E_0 =\omega /2$ and of the $\vert \psi_0(0) \vert^2=\Big({\omega \over \pi}\Big)^{1/2}$.

\end{document}